\def\Im{{\text{Im}}\,}

\def\pF{p_{\text{F}}}
\def\vF{v_{\text{F}}}
\def\NF{N_{\text{F}}}

\def\epsilonF{\epsilon_{\text F}}

\def\sgn{{\text{sgn\,}}}
\def\be{\begin{equation}}
\def\ee{\end{equation}}
\def\bea{\begin{eqnarray}}
\def\eea{\end{eqnarray}}
\def\bse{\begin{subequations}}
\def\ese{\end{subequations}}

\documentclass[prb,twocolumn,amsmath,amssymb,eqsecnum,preprintnumbers]{revtex4}
\usepackage{graphicx}% Include figure files
\usepackage{dcolumn}% Align table columns on decimal point
\usepackage{bm}% bold math
\usepackage{bbm}
\usepackage{color}
\usepackage{soul}
\usepackage{enumitem}
\begin{document}
%\preprint{KITP Preprint NSF-KITP-05-xx}
%\preprint{Phys. Rev. B {\bf 84}, 134401 (2014)}
\title{Soft modes in Fermi liquids at arbitrary temperatures}
\author{D. Belitz$^{1,2}$, and T.R. Kirkpatrick$^2$}
\affiliation{$^{1}$ Department of Physics and Institute for Fundamental Science,
                    University of Oregon, Eugene, OR 97403, USA\\
                   $^{2}$ Materials Science Institute, University of Oregon, Eugene,
                    OR 97403, USA\\
                  $^{2}$ Institute for Physical Science and Technology,
                    University of Maryland, College Park,
                    MD 20742, USA
            }
\date{\today}

\begin{abstract}
We use kinetic-theory methods to analyze Landau Fermi-liquid theory, and in particular to investigate the number 
and nature of soft modes in Fermi liquids, both in the hydrodynamic and the collisionless regimes. In the 
hydrodynamic regime we show that Fermi-liquid theory is consistent with Navier-Stokes hydrodynamics at all temperatures.
The soft modes are the ones familiar from classical hydrodynamics that are controlled by the five conservation
laws; namely, two first-sound modes, two shear diffusion modes, and one heat diffusion mode. These
modes have a particle-like spectrum and are soft, or scale invariant, at all temperatures. In the 
collisionless regime we show that the entire single-particle distribution function is soft with a continuous part of the spectrum. 
This continuous soft mode, which is well known but often not emphasized, has important physical 
consequences, e.g., for certain quantum phase transitions. In addition, there are the well known soft 
zero-sound excitations that describe angular fluctuations of the Fermi surface; their spectra are particle-like.
They are unrelated to conservation laws, acquire a mass at any nonzero temperature, and their number 
depends on the strength of the quasiparticle interaction. We also discuss the fates of these two families
of soft modes as the temperature changes. With increasing temperature the size of the collisionless regime 
shrinks, the damping of the modes grows, and eventually all of the collisionless modes become overdamped.
In their stead the five hydrodynamic modes appear in the hydrodynamic regime at asymptotically
low frequencies. The two families of soft modes are unrelated and have very different physical origins.
In charged Fermi liquids the first-sound modes in the hydrodynamic regime and the $\ell=0$ zero-sound 
modes in the collisionless regime get replaced by plasmons, all other modes remain soft.
\end{abstract}

%\pacs{7B.2.Di; 7B.7.Lh; 75.30.Ds}

\maketitle

\section{Introduction}
\label{sec:I}

Fluid mechanics is one of the oldest and most important subfields of physics, going back to Euler's work in the mid-1700s.
Classical fluid mechanics\cite{Landau_Lifshitz_VI_1959} demonstrated that by applying the principles of Newtonian 
mechanics one can successfully describe the behavior of strongly interacting condensed many-particle systems. 
The advent of quantum mechanics raised the question of how to describe fluids of quantum particles at low 
temperatures $T$, in particular fermionic ones such as He-3 or the electron fluid in a metal. 
Landau's Fermi-liquid theory and its generalization to charged systems provided an answer by combining kinetic
theory with a phenomenological description of strong interactions and the notion of `quasiparticles'; fictitious 
entities that allow for a mapping of the spectrum of the strongly interacting system onto that of a noninteracting 
Fermi gas.\cite{Landau_Lifshitz_IX_1991, Baym_Pethick_1991, Pines_Nozieres_1989} 

\begin{figure}[b]
\vskip 0pt
\includegraphics[width=8cm]{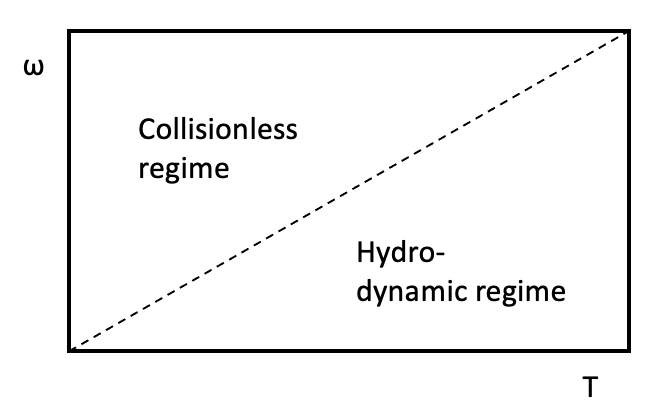}
\caption{Schematic locations of the collisionless and hydrodynamic regimes in the frequency-temperature plane.
              The boundary (dashed line) is not sharp, and its functional form depends on the temperature dependence
              of the relevant scattering time.}
\label{fig:1}
\end{figure}

LFL theory was initially applied to Fermi liquids\cite{semantics_footnote} at zero temperature in the absence of impurities, 
where it predicted, {\it inter alia}, the
collective modes known as zero sound. Like ordinary, or first, sound they are soft or massless modes, i.e., their frequency goes
to zero linearly with the wave number. In other words, these excitations are scale invariant, in contrast to massive or gapped
ones whose frequency approaches a nonzero constant value as the wave number goes to zero. Unlike first sound, their 
existence has nothing to do with conservation laws. In fact, any
nonzero temperature makes the zero-sound modes massive, whereas the masslessness of first sound is protected by the
conservation of particle number and momentum. More generally, one needs to distinguish between the collisionless regime
at low temperatures, where the streaming term in a kinetic equation dominates over the collision integral, and the hydrodynamic
regime at higher temperatures, where the collisions dominate. The former is usually delineated by the requirement $\omega\tau\gg 1$,
with $\omega$ the frequency of an excitation and $\tau$ a relevant scattering time, and the latter by $\omega\tau\ll 1$. Since
$\tau(T\to 0)\to\infty$, the collisionless regime at $T=0$ extends all the way to zero frequency, see Fig.~\ref{fig:1}.

These considerations lead to three questions: (1) What is the relation between LFL theory and classical hydrodynamics in
the hydrodynamic regime? Since the soft modes in the hydrodynamic regime of both classical fluids and quantum fluids are
governed by conservation laws, the structure of the theories should be the same as far as the soft modes are concerned,
yet LFL theory is usually assumed to be valid only at low (compared to the Fermi temperature) temperatures. (2) What is the
relation, if any, between the soft modes in the hydrodynamic and collisionless regimes? The experimentally observed transition
from first sound to zero sound in He-3, with a region of strong damping in between (see, e.g., Ref.~\onlinecite{Baym_Pethick_1991}),
is often referred to as a `crossover', but the very different origins of the excitations make it implausible that they are continuously
connected in any physical sense. This is part of a more general question: What happens to classical soft modes with decreasing
temperature, and to quantum soft modes with increasing temperature? (3) What is the relation, if any, between the collective 
zero-sound modes that are particle-like
in the sense that they are characterized by a well-defined frequency with a small damping coefficient, and the continuous
excitation known, in a LFL context, as the particle-hole continuum,\cite{Pines_Nozieres_1989} that is NOT particle-like
but nonetheless scale invariant with the frequency scaling linearly with the wave number?

The purpose of this paper is to discuss and answer these questions. With respect to question (1), we will show that LFL
theory describes the soft modes of a Fermi system qualitatively correctly for all temperatures. In particular, it yields five soft modes
in the hydrodynamic regime, namely, two (first) sound modes, two shear diffusion modes, and one heat diffusion mode,
that are completely consistent with the linearized Navier-Stokes equations. With respect to question (2), we will discuss
that there is no connection between the soft modes in the two regimes. With decreasing temperature the hydrodynamic
regime shrinks and the hydrodynamic modes become confined to a smaller and smaller region of parameter space and
finally disappear. In their stead, a new family of soft modes emerges in the collisionless regime that is governed by
different physics. With respect to question (3), we will show that the particle-like zero-sound excitations and the 
continuum mode (which we will refer to as the `unparticle' mode, using a term coined in Ref.~\onlinecite{Georgi_2007})
are part of one and the same scale invariant spectrum, and hence just different aspects of the same excitation. While
historically the zero-sound modes have received much more attention, the continuum unparticle mode has important
physical consequences. A particularly striking one is that it is responsible for driving the quantum ferromagnetic
transition generically first order by coupling to the magnetic order parameter.\cite{Brando_et_al_2016a}

Experimental studies of quantum hydrodynamics used to be limited since there was only one known
neutral Fermi liquid, He-3, with very limited opportunities for tuning the interaction parameters, while in metals,
which represent charged Fermi liquids, the hydrodynamic regime tends to be dominated by impurity scattering,
which drastically alters the hydrodynamic behavior from that of clean systems. This has changed in recent years
due to the availability of cold-atom systems that allow for the realization of Fermi liquids with widely tunable
parameters.\cite{Cao_et_al_2011, Bloch_Dalibard_Nascimbene_2012, Joseph_Elliott_Thomas_2015, Bluhm_Hou_Schaefer_2017, 
Patel_et_al_2020, Wang_et_al_2021} This is a further motivation for studying quantum hydrodynamics in more detail.

In this paper we will use the usual Boltzmann/Landau kinetic equation\cite{Landau_Lifshitz_IX_1991,Landau_Lifshitz_X_1981}
and construct explicit solutions that identify the soft modes in the system.
In a second paper (Ref.~\onlinecite{Kirkpatrick_Belitz_2022}, to be referred to as Paper II) we add a Langevin force
to the quantum kinetic equation, and in particular derive fluctuating Navier-Stokes equations for  a fermionic
quantum fluid.

This paper is organized as follows. In Sec.~\ref{sec:II} we briefly review key elements of LFL theory and formulate
the kinetic equation that governs the time evolution of the single-particle distribution function. In Sec.~\ref{sec:III}
we discuss the solution of the linearized kinetic equation in the hydrodynamic regime, and show that the results
for the soft modes are the same as in the theory of classical fluids. In Sec.~\ref{sec:IV} we discuss the solution
in the collisionless regime, with some emphasis on the continuum unparticle excitation, and highlight the
fundamental differences between the soft modes in the two regimes. We conclude in Sec.~\ref{sec:V} with
a discussion of various aspects of the theory and its implications. Some technical details regarding thermodynamic
relations, continuity equations, and collision operators, are relegated to three appendices.

\section{Kinetic Theory for Fermions}
\label{sec:II}

In this section we recall general aspects of Landau Fermi Liquid (LFL) theory\cite{Landau_Lifshitz_IX_1991, Baym_Pethick_1991} 
and discuss the generalized Boltzmann equation that governs the time evolution of the distribution function.
For our detailed discussion we will consider %Fermi liquids that consist of interacting fermions only. 
systems where the fermions interact only with each other.
However,
the formalism can easily be generalized to include the interaction of fermions with impurities, or with bosons
such as phonons or magnons. Some relevant linearized collision operators are given in Appendix~\ref{app:C}.

\subsection{Landau's Fermi-Liquid theory}
\label{subsec:II.A}

Consider a description of a Fermi liquid in terms of Landau quasiparticles 
(QPs),  and let $f({\bm p},{\bm x},t)$ be the single-QP
distribution function that depends on the momentum ${\bm p}$, the real-space position
${\bm x}$, and the time $t$. For simplicity we consider spinless fermions; the theory can easily be
generalized to include spin. If $e({\bm x},t)$ is the energy density of the system as a whole, then the QP
energy $\epsilon({\bm p},{\bm x},t)$ is defined via the change $\delta e$ of the energy density under a change 
$\delta f$ of the distribution function,
\bse
\label{eqs:2.1}
\be
\delta e({\bm x},t) = \frac{1}{V}\sum_{\bm p} \epsilon({\bm p},{\bm x},t)\,\delta f({\bm p},{\bm x},t)
\label{eq:2.1a}
\ee
with $V$ the system volume.\cite{thermodynamic_limit_footnote} For later reference we also introduce %fluctuations 
variations
of the number density
\be
\delta n({\bm x},t) = \frac{1}{V}\sum_{\bm p} \delta f({\bm p},{\bm x},t)\ ,
\label{eq:2.1b}
\ee
and of the fluid velocity
\be
\delta {\bm u}({\bm x},t) = \frac{1}{nm} \frac{1}{V}\sum_{\bm p} {\bm p}\,\delta f({\bm p},{\bm x},t)\ .
\label{eq:2.1c}
\ee
\ese
with $m$ the bare fermion mass and $n$ the equilibrium number density, see Eq.~(\ref{eq:2.5}) below.

In equilibrium, $\epsilon({\bm p},{\bm x},t)$ is independent of space and time. For simplicity, we assume an isotropic Fermi liquid, 
so $\epsilon_{\text{eq}}({\bm p})$ depends on $p = \vert{\bm p}\vert$ only, and we write
$\epsilon_{\text{eq}}({\bm p}) \equiv \epsilon_p$.  The QP velocity is
\be
{\bm v}_p = {\bm\nabla}_{\bm p}\,\epsilon_p\ .
\label{eq:2.2}
\ee
Near the Fermi surface one has $\epsilon_p = \mu + \vF^*(p - \pF)$, with $\mu$ the chemical
potential, $\pF$ the Fermi wave number, and $\vF^* = \pF/m^*$ the Fermi velocity with $m^*$ the QP effective mass.
Consequently, for ${\bm p}$ on the Fermi surface one has
${\bm v}_p = \vF^*\,{\hat {\bm p}} = {\bm p}/m^*$. (We use units such that $\hbar = k_{\text B} = 1$.)
At zero temperature, $\mu(T=0)  = \pF^2/2m^* \equiv \epsilonF$. 

%\tcr{\st{Within LFL theory}} 
The equilibrium distribution function is given by the Fermi-Dirac distribution 
\bse
\label{eqs:2.3}
\be
f_{\text{eq}}({\bm p}) = 1/(\exp(\xi_p/T) + 1)\ , 
\label{eq:2.3a}
\ee
with $\xi_p = \epsilon_p-\mu$. For later reference we also define
\bea
w({\bm p}) &=& -\partial f_{\text{eq}}({\bm p})/\partial\epsilon_p = \frac{1}{T}\,f_{\text{eq}}({\bm p})\left[1 - f_{\text{eq}}({\bm p})\right] 
\nonumber\\
&=& \frac{1}{4T \cosh^2(\xi_p/2T)}\ ,
\label{eq:2.3b}
\eea
\ese
which plays the role of a weight function. Note that $f_{\text{eq}}$ and $w$ depend on the modulus of ${\bm p}$ only. 
It is useful to define a scalar product $\langle\ldots\vert\ldots\rangle$ of ${\bm p}\,$-dependent functions
\bse
\label{eqs:2.4}
\be
\left\langle g({\bm p})\vert h({\bm p})\right\rangle = \frac{1}{V}\sum_{\bm p} w({\bm p})\,g({\bm p})\,h({\bm p})\ ,
\label{eq:2.4a}
\ee
and an average with respect to the weight function $w$ by
\be
\langle g({\bm p})\rangle_w = \left\langle g({\bm p})\vert 1\right\rangle/N_0
\label{eq:2.4b}
\ee
where
\be
N_0 = \langle 1\vert 1\rangle = \frac{1}{V}\sum_{\bm p} w({\bm p})
\label{eq:2.4c}
\ee
\ese
normalizes the weight function. 
For $T\to 0$, $N_0 = \NF^* + O(T^2)$, where $\NF^* = \pF m^*/2\pi^2$ is the QP density of states at the Fermi surface

The equilibrium number density is
\be
n = \frac{N}{V} = \frac{1}{V} \sum_{\bm p} f_{\text{eq}}({\bm p}) \ ,
\label{eq:2.5}
\ee
and the physical mass density is $\rho = nm$. At $T=0$, $n = \pF^3/6\pi^2$.
The internal energy in equilibrium we will denote by $E$, and the equilibrium energy density by $e = E/V$.

%Away from equilibrium, $\epsilon$ is a complicated functional of $f$. 
Within LFL theory, a change $\delta f$ of the distribution function leads to a change of 
$\epsilon$ given by
\be
\delta\epsilon({\bm p},{\bm x},t) = \frac{1}{N_0 V}\sum_{{\bm p}'} F({\bm p},{\bm p}')\,\delta f({\bm p}',{\bm x},t)\ .
\label{eq:2.6}
\ee
Here $F({\bm p},{\bm p}') = F({\bm p}',{\bm p})$ is a function that parameterizes the interaction between the QPs, and 
the factor of $N_0$ serves to make $F$ dimensionless. The interaction function $F$ is the kernel of the integral
equation obeyed by the QP velocity,\cite{Landau_Lifshitz_IX_1991}
\be
{\bm v}_p =  \frac{1}{m}\,{\bm p} -  \frac{1}{N_0 V}\sum_{{\bm p}'} w({\bm p}')\,F({\bm p},{\bm p}')\,{\bm v}_{p'}\ .
\label{eq:2.7}
\ee
%For ${\bm p}$ and ${\bm p}'$ on the Fermi surface, $F$ can be expanded in Legendre polynomials,
For the discussion in Sec.~\ref{sec:III} we keep only the $\ell=0$ and $\ell=1$ terms in an angular-momentum expansion and write 
\be
%F({\bm p},{\bm p}') =  \sum_{\ell=0}^{\infty} F_{\ell}\, P_{\ell}({\hat{\bm p}}\cdot{\hat{\bm p}}')\ .
F({\bm p},{\bm p}') = F_0 + F_1\,\frac{ {\bm p}\cdot{\bm p}'}{\langle {\bm p}^2 \rangle_w}\ .
 \label{eq:2.8}
\ee
At $T=0$ this reduces to the usual definition of the Landau parameter $F_1$, which determines the relation between the 
QP effective mass $m^*$ and the bare fermion mass $m$ 
 via\cite{Landau_Lifshitz_IX_1991, Baym_Pethick_1991}
 \be
 m^* = m(1 + F_1/3)\ .
 \label{eq:2.9}
 \ee
Note that Eq.~(\ref{eq:2.9}), with $F_1$ as defined in Eqs.~(\ref{eq:2.8}), remains valid if one performs its
usual derivation\cite{Baym_Pethick_1991} at nonzero temperature. With the model interaction given by
Eq.~(\ref{eq:2.8}) the integral equation (\ref{eq:2.7}) is separable and can be solved. We find
\bse
\label{eqs:2.10}
\be
{\bm v}_p = {\bm p}/m^*\ ,
\label{eq:2.10a}
\ee
and hence
\be
\epsilon_p = \frac{1}{2}\,{\bm p}\cdot{\bm v}_p = p^2/2m^*\ .
\label{eq:2.10b}
\ee
\ese
Here we have used Eqs.~(\ref{eq:A.12}) and (\ref{eq:A.19b}), and we ignore a $p\,$-independent contribution to $\epsilon_p$ that also
depends on the FL interaction.\cite{Fetter_Walecka_1971} Note that within this model this holds in general, 
whereas for a general interaction function $F$ it holds only for ${\bm p}$ on the Fermi surface.\cite{v_p_footnote}
%Within LFL theory, the Fermi-liquid parameters $F_{\ell}$ completely 
%characterize the QP interaction. 
%The parameter $F_0$ is of particular significance. 
The Landau parameter $F_0$ is a positive number for a neutral Fermi liquid with a 
short-ranged (SR) repulsive interaction between the QPs. However, for
 a charged Fermi liquid (such as the conduction-electron system in a metal) $F_0$ must be augmented by the 
 long-ranged (LR) Coulomb interaction:\cite{Silin_1957, Pines_Nozieres_1989}
 \be
 F_0 \to F_0 + 4\pi N_0\, e^2/{\bm k}^2\qquad \text{(LR case)}\ ,
 \label{eq:2.11}
 \ee
where we have anticipated a spatial Fourier transform from the real-space position variable ${\bm x}$ to a
wave vector ${\bm k}$.
 
The kinetic equation that governs the time evolution of the distribution function $f$ is
\be
\frac{d}{dt}\,f({\bm p},{\bm x},t) = \left(\frac{\partial f}{\partial t} \right)_{\text{coll}}\ .
\label{eq:2.12}
\ee
This holds in complete generality: The total time derivative of $f$ on the left-hand side 
equals the collision integral on the right-hand side, i.e., the temporal change of $f$ due to collisions between 
the QPs. The left-hand side consists of three terms:
\bse
\label{eqs:2.13}
\be
\frac{d}{dt} f = \partial_t f + {\bm\nabla}_{\bm x} f\cdot{\dot{\bm x}} + {\bm\nabla}_{\bm p} f \cdot{\dot{\bm p}}\ .
\label{eq:2.13a}
\ee
The velocity is given by
\be
{\dot{\bm x}} = {\bm\nabla}_{\bm p}\, \epsilon({\bm p},{\bm x},t) \ ,
\label{eq:2.13b}
\ee
and ${\dot{\bm p}}$ is equal to the force due to the spatially inhomogeneous energy, i.e., the potential energy of the
QP interaction,
\be
{\dot{\bm p}} = - {\bm\nabla}_{\bm x}\,\epsilon({\bm p},{\bm x},t) = - {\bm\nabla}_{\bm x}\,\delta\epsilon({\bm p},{\bm x},t)\ .
\label{eq:2.13c}
\ee
To linear order in small deviations $\delta f$ from the equlibrium distribution we thus have
\be
\frac{d}{dt} f = \partial_t\, \delta f + {\bm v}_p\cdot{\bm\nabla}_{\bm x}\, \delta f 
                        - {\bm v}_p\cdot \frac{\partial f_{\text{eq}}}{\partial\epsilon_p}\,{\bm\nabla}_{\bm x}\,\delta\epsilon\ .
\label{eq:2.13d}
\ee
\ese
with $\delta\epsilon$ given by Eq.~(\ref{eq:2.6}). In what follows it will be convenient to write
\be
\delta f({\bm p},{\bm x},t) = w({\bm p})\,\phi({\bm p},{\bm x},t) 
\label{eq:2.14}
\ee
with $w({\bm p})$ from Eq.~(\ref{eq:2.3b}).

\subsection{The linearized kinetic equation}
\label{subsec:II.B}

We write the collision integral as a linear collision operator $\Lambda$ acting on $\delta f$:
\be
\left(\frac{\partial f}{\partial t} \right)_{\text{coll}} = \Lambda({\bm p})\,\delta f({\bm p},{\bm x},t)
\label{eq:2.15}
\ee
After a temporal Laplace transform with $z$ as the complex frequency\cite{Laplace_trafo_footnote} and a spatial Fourier transform with ${\bm k}$ 
as the wave vector the kinetic equation (\ref{eq:2.12}) takes the form  
\bse
\label{eqs:2.16}
\be
\left[-i z + L_{\bm k}({\bm p})\right] \phi({\bm p},{\bm k},z) = \phi({\bm p},{\bm k},t=0)\ .
\label{eq:2.16a}
\ee
Here $L_{\bm k}({\bm p})$ is a linearized kinetic operator given by
\bea
L_{\bm k}({\bm p}) &=& i\,{\bm k}\cdot{\bm v}_p\,\left[1 + \frac{1}{N_0 V} \sum_{{\bm p}'} w({\bm p}')\,F({\bm p}',{\bm p})\,R_{{\bm p}\to{\bm p}'}\right] 
\nonumber\\
&& - \Lambda({\bm p})\ ,
\label{eq:2.16b}
\eea
\ese
with the replacement operator $R$ defined by $R_{{\bm p}\to{\bm p}'} g({\bm p}) = g({\bm p}')$ for any ${\bm p}$-dependent function $g$.  
The first contribution to $L_{\bm k}({\bm p})$ represents the streaming term and the QP interaction term in Eq.~(\ref{eq:2.13d}). The collision
operator $\Lambda$ must respect the five collision invariants that are unaffected by QP collisions,
viz., the density, the three components of the momentum, and the energy.\cite{conservation_laws_footnote}
%We define a scalar product $(..\vert ..)$
%in the space of momentum-dependent functions by
%\bse
%\label{eqs:2.15}
%\be
%\left\langle g({\bm p})\vert h({\bm p})\right\rangle = \frac{1}{V}\sum_{\bm p} w({\bm p})\,g({\bm p})\,h({\bm p})\ ,
%\label{eq:2.15a}
%\ee
%and an average with respect to the weight function $w$ by
%\be
%\langle g({\bm p})\rangle_w = \left\langle g\vert 1\right\rangle/N_0
%\label{eq:2.15b}
%\ee
%\ese
%with $N_0 = \langle 1\vert 1\rangle$ from Eq.~(\ref{eq:2.3c}). 

Using the scalar product defined in Eq.~(\ref{eq:2.4a}) we can write the density, velocity, 
and energy fluctuations from Eqs.~(\ref{eqs:2.1}) as
\bse
\label{eqs:2.17}
\bea
\delta n({\bm x},t) &=& \left\langle1\vert\phi({\bm p},{\bm x},t)\right\rangle\ ,
\label{eq:2.17a}\\
\delta{\bm u}({\bm x},t) &=& \frac{1}{nm} \left\langle{\bm p}\vert\phi({\bm p},{\bm x},t)\right\rangle\ ,
\label{eq:2.17b}\\
\delta e({\bm x},t) &=& \left\langle \epsilon_p \vert \phi({\bm p},{\bm x},t)\right\rangle\ .
\label{eq:2.17c}
\eea
\ese
We separate the momentum into a longitudinal (with respect to the wave vector ${\bm k}$) and two transverse components:
\be
{\bm p} = ({\hat{\bm k}}\cdot{\bm p}){\hat{\bm k}} + \sum_{i=1,2} ({\hat{\bm k}}_{\perp}^{(i)} \cdot {\bm p}) {\hat{\bm k}}_{\perp}^{(i)}
\label{eq:2.18}
\ee
with ${\hat{\bm k}}$ the unit vector in ${\bm k}$-direction, and two unit vectors ${\hat{\bm k}}_{\perp}^{(1,2)}$ that are perpendicular to
${\bm k}$ and to each other. This separates the velocity fluctuations into one longitudinal component $\delta u_L$ and two
transverse components $\delta u_{\perp}^{(1,2)}$ and we write the five fluctuations from Eqs.~(\ref{eqs:2.17}) in the form
\bse
\label{eqs:2.19}
\bea
\delta n({\bm x},t) &=& \langle a_1({\bm p})\vert\phi({\bm p},{\bm x},t)\rangle\ ,
\label{eq:2.19a}\\
\delta u_L({\bm x},t) &=& \frac{1}{nm} \langle a_2({\bm p})\vert\phi({\bm p},{\bm x},t)\rangle\ ,
\label{eq:2.19b}\\
\delta u_{\perp}^{(1)}({\bm x},t) &=& \frac{1}{nm} \langle a_3({\bm p})\vert\phi({\bm p},{\bm x},t)\rangle\ ,
\label{eq:2.19c}\\
\delta u_{\perp}^{(2)}({\bm x},t) &=& \frac{1}{nm}\langle a_4({\bm p})\vert\phi({\bm p},{\bm x},t)\rangle\ ,
\label{eq:2.19d}\\
\delta e({\bm x},t) &=& \langle a_5({\bm p})\vert\phi({\bm p},{\bm x},t)\rangle\  + \langle\epsilon_p\rangle_w\, \delta n({\bm x},t)\ .
\nonumber\\
\label{eq:2.19e}
\eea
Also of interest are the temperature fluctuations
\be
\delta T({\bm x},t) = \frac{1}{c_V} \langle a_5({\bm p})\vert\phi({\bm p},{\bm x},t)\rangle\ ,
\label{eq:2.19f}
\ee
the fluctuations the entropy density $s=S/V$,
\be
T \delta s({\bm x},t) = \delta e({\bm x},t) - \mu\,\delta n({\bm x},t)\, 
\label{eq:2.19g}
\ee
and the pressure fluctuations
\be
\delta p({\bm x},t) = \left(\frac{\partial p}{\partial T}\right)_{N,V} \delta T({\bm x},t) + \left(\frac{\partial p}{\partial n}\right)_{T,V} \delta n({\bm x},t)\ .
\label{eq:2.19h}
\ee
\ese
Here we have defined
\bse
\label{eqs:2.20}
\bea
a_1({\bm p}) &=& 1\ ,
\label{eq:2.20a}\\
a_2({\bm p}) &=&  {\hat{\bm k}}\cdot{\bm p}\ ,
\label{eq:2.20b}\\
a_3({\bm p}) &=& {\hat{\bm k}}_{\perp}^{(1)} \cdot{\bm p}\ ,
\label{eq:2.20c}\\
a_4({\bm p}) &=& {\hat{\bm k}}_{\perp}^{(2)} \cdot{\bm p}\ ,
\label{eq:2.20d}\\
a_5({\bm p}) &=& \epsilon_p - \langle\epsilon_p\rangle_w\ ,
\label{eq:2.20e}
\eea
\ese
and we have split the energy fluctuation into its overlap with the density fluctuation plus a part $\langle a_5\vert\phi\rangle$ 
that is orthogonal to the density. We will also need the normalizations of these vectors. We define
\bse
\label{eqs:2.21}
\be
A_{\alpha}^{-2} = \langle a_{\alpha}({\bm p}) \vert a_{\alpha}({\bm p})\rangle
\label{eq:2.21a}
\ee
and find
\bea
A_1^{-2} &=& \langle 1\vert 1\rangle = (1+F_0)(\partial n/\partial\mu)_T \ ,
\label{eq:2.21b}\\
A_2^{-2} &=& A_3^{-2} = A_4^{-2} = \frac{1}{3}\,\langle{\bm p}\vert{\bm p}\rangle = n m^*\ , 
\label{eq:2.21c}\\
A_5^{-2} &=&\left\langle a_5({\bm p})\vert a_5({\bm p})\right\rangle = T c_V\ ,
\label{eq:2.21d}
\eea
\ese
with $c_V$ the specific heat at constant volume.
See Appendix~\ref{app:A.1} for the final equalities in Eqs.~(\ref{eqs:2.21}), and Appendix~\ref{app:A.2} for the 
origin of Eq.~(\ref{eq:2.19f}). Equation~(\ref{eq:2.19g}) follows from Eq.~(\ref{eq:A.11}).

Now consider the five conservation laws. Particle number conservation implies
\bea
0 &=& \frac{d}{dt}\,\delta n({\bm x},t) = \frac{1}{V} \sum_{\bm p} w({\bm p})\,\frac{d}{dt}\,\phi({\bm p},{\bm x},t) 
\nonumber\\
&=&  \frac{1}{V} \sum_{\bm p} w({\bm p})\,\Lambda({\bm p})\,\phi({\bm p},{\bm x},t)
   = \left\langle1\vert\Lambda({\bm p})\vert\phi({\bm p},{\bm x},t)\right\rangle\ .
\nonumber\\
\label{eq:2.22}
\eea
Consequently, $\Lambda({\bm p})$ has a zero eigenvalue with eigenvector $a_1({\bm p})$.
Analogously, the remaining four conservation laws are reflected by four additional zero eigenvalues.
We will refer to the space spanned by the $a_{\alpha}({\bm p})$ listed in Eqs.~(\ref{eqs:2.20}) as the
`zero-eigenvector space' and denote it by ${\cal L}_0$. The space orthogonal to ${\cal L}_0$ we will
refer to as the `orthogonal space' and denote it by ${\cal L}_{\perp}$. 
For a discussion of the continuity equations related to the five conservations laws see Appendix~\ref{app:B}.

These properties of the collision operator suffice for determining the soft modes in the hydrodynamic regime. 
In the collisionless regime, the collision operator can be neglected to leading order. If desirable, one can
construct explicit model collision operators that have the required five zero eigenvalues, see Sec.~\ref{sec:IV} 
and Appendix~\ref{app:C}.

The solution of the kinetic equation (\ref{eq:2.16a}) is qualitatively different depending on whether 
%$\vF k\tau \ll 1$ (``hydrodynamic regime") or $\vF k \tau \gg 1$ (``collisionless regime"). Here $k = \vert{\bm k}\vert$. 
the collision operator or the streaming and interaction terms in the kinetic operator $L_{\bm k}({\bm p})$ dominate. 
The corresponding regions of parameter space are known as the ``hydrodynamic regime'' and the ``collisionless regime",
respectively, see Fig.~\ref{fig:1}. We will discuss the hydrodynamic regime first, and the collisionless regime second.

\section{Solutions of the kinetic equation I: Hydrodynamic regime}
\label{sec:III}

The system is always in the hydrodynamic regime for fixed nonzero temperature at asymptotically small wave numbers
(or frequencies), or for fixed wave number or frequency at sufficiently high temperature. This is because of the linear wave-number dependence of 
the streaming/interaction contribution to $L_{\bm k}({\bm p})$ and the fact that collisions become less frequent 
with decreasing temperature. 

\subsection{Short-ranged case}
\label{subsec:III.A}

In the hydrodynamic regime, the collision operator dominates over the first term on the right-hand side of Eq.~(\ref{eq:2.16b}).
As a result, all modes are massive (i.e., they have a nonzero frequency at ${\bm k}=0$) except for the five hydrodynamic modes
tied to the five conservation laws. The corresponding hydrodynamic frequencies are given by the eigenvalues of the 
kinetic operator $L_{\bm k}({\bm p})$ defined in Eq.~(\ref{eq:2.16b}), and the hydrodynamic modes are given by the corresponding
left eigenfunctions, see Eqs.~(\ref{eqs:2.19}). Both the eigenvalues and the eigenfunctions can be determined by 
studying the perturbed zero eigenvalues.\cite{Dorfman_vanBeijeren_Kirkpatrick_2021}
We thus consider the left eigenproblem
\be
\langle \psi_{\alpha}^{\text{L}}({\bm k},{\bm p}) \vert \, L_{\bm k}({\bm p})\, = \langle \psi_{\alpha}^{\text{L}}({\bm k},{\bm p})\vert\, \omega_{\alpha}({\bm k})\ .
\label{eq:3.1}
\ee
We are interested in the small-${\bm k}$ behavior of the five eigenvalues that are zero at ${\bm k}=0$. Accordingly, we
perform an expansion in powers of ${\bm k}$:
\bse
\label{eqs:3.2}
\be
\omega_{\alpha}({\bm k}) = \omega_{\alpha}^{(1)}({\bm k}) + \omega_{\alpha}^{(2)}({\bm k}) + O(k^3)\qquad (\alpha = 1,2,3,4,5)
\label{eq:3.2a}
\ee
where $\omega_{\alpha}^{(n)} = O(k^n)$. Analogously, we expand the eigenfunctions
\bea
\langle \psi_{\alpha}^{\text{L}}({\bm k},{\bm p}) \vert &=& \langle \psi_{\alpha}^{\text{L}(0)}({\bm k},{\bm p}) \vert 
   + \langle \psi_{\alpha}^{\text{L}(1)}({\bm k},{\bm p})\vert  
      + \langle \psi_{\alpha}^{\text{L}(2)}({\bm k},{\bm p})\vert
\nonumber\\
 && + O(k^3)\ .
\label{eq:3.2b}
\eea
\ese
$\psi_{\alpha}^{\text{L}(0)}$ can depend on ${\bm k}$ at most via $\hat{\bm k}$, $\psi_{\alpha}^{\text{L}(1)}$ is linear
in $k$, etc. For notational simplicity we will not show this ${\bm k}$-dependence explicitly from here on and write
$\psi_{\alpha}^{\text{L}(0)}({\bm k},{\bm p}) \equiv \psi_{\alpha}^{\text{L}(0)}({\bm p})$, etc.
In the same expansion, the linearized kinetic operator is (see Eq.~(\ref{eq:2.16b}))
\be
L_{\bm k}({\bm p}) = -\Lambda({\bm p}) + L_{\bm k}^{(1)}({\bm p})\ .
\label{eq:3.3}
\ee
The operator $L_{\bm k}^{(1)}({\bm p})$ that is linear in ${\bm k}$ has two contributions:
\bse
\label{eqs:3.4}
\be
 L_{\bm k}^{(1)}({\bm p}) =  L_{\bm k}^{(1,1)}({\bm p}) +  L_{\bm k}^{(1,2)}({\bm p}) 
 \label{eq:3.4a}
 \ee
 where
 \be
L_{\bm k}^{(1,1)}({\bm p}) = i\,{\bm k}\cdot{\bm v}_p
\label{eq:3.4b}
\ee
reflects the streaming term for noninteracting QPs with mass $m^*$, and
\be
 L_{\bm k}^{(1,2)}({\bm p}) = i\,{\bm k}\cdot{\bm v}_p\, \frac{1}{N_0 V}\sum_{{\bm p}'} w({\bm p}')\,F({\bm p}',{\bm p})\,R_{{\bm p}\to{\bm p}'} 
 \label{eq;3.4c}
 \ee
 reflects the QP interaction. If we use the LFL model interaction from Eq.~(\ref{eq:2.8}), then by utilizing
 the scalar product notation this can be written as
 \be
 L_{\bm k}^{(1,2)}({\bm p}) = \frac{F_0}{\langle 1\vert 1\rangle}\,\big\vert i{\bm k}\cdot{\bm v}_p\big\rangle \big\langle 1\big\vert
       + \frac{F_1}{\langle{\bm p}\vert{\bm p}\rangle}\,\big\vert(i{\bm k}\cdot{\bm v}_p){\bm p}\big\rangle \cdot \big\langle{\bm p}\big\vert\ .
\label{eq:3.4d} 
\ee
 \ese
As we pointed out in Sec.~\ref{sec:II}, this model for the interaction implies the simple form (\ref{eq:2.10a}) for the QP velocity.
That is, within this model ${\bm v}_p \in {\cal L}_0$. We will nonetheless usually write ${\bm v}_p$ rather than ${\bm p}/m^*$
in order to point out several aspects of the theory that will emerge if one considers a more general QP interaction function
so that ${\bm v}_p$ has a component in the orthogonal space ${\cal L}_{\perp}$.
 
To zeroth order in this wave-number expansion we have
\be
\big\langle \psi_{\alpha}^{\text{L}(0)}({\bm p}) \big\vert\,\Lambda({\bm p}) = 0\ ,
\label{eq:3.5}
\ee
which implies that the $\psi_{\alpha}^{\text{L}(0)}({\bm p})$ are linear combinations of the $a_{\alpha}({\bm p})$ from Eqs.~(\ref{eqs:2.20}):
\be
\langle \psi_{\alpha}^{\text{L}(0)}({\bm p}) \vert = \sum_{\beta=1}^5 c_{\beta}^{(\alpha)}(\hat{\bm k})\,\langle a_{\beta}({\bm p})\vert\ .
\label{eq:3.6}
\ee
By systematically going to higher order in $k$ we can now determine the hydrodynamic modes and their eigenvalues.
In particular, in order to determine the coefficients $c_{\beta}^{(\alpha)}$ in Eq.~(\ref{eq:3.6}) one needs to go to $O(k)$.

\subsubsection{Speed of first sound}
\label{subsubsec:III.A.1}

To first order in the $k$-expansion we have
\be
\langle \psi_{\alpha}^{\text{L}(0)}({\bm p})\vert\, L_{\bm k}^{(1)}({\bm p})\, - \langle \psi_{\alpha}^{\text{L}(1)}({\bm p}) \vert\, \Lambda({\bm p})\, 
%\nonumber\\
= \omega_{\alpha}^{(1)}({\bm k})\,\langle\psi_{\alpha}^{\text{L}(0)}({\bm p})\vert\ .
\label{eq:3.7}
\ee
Multiplying from the right with $w({\bm p})\,a_{\gamma}({\bm p})$ and summing over ${\bm p}$ we obtain
\bse
\label{eqs:3.8}
\be
\sum_{\beta} c_{\beta}^{(\alpha)}(\hat{\bm k}) \left( L^{(1)}_{\beta\gamma}({\bm k},{\bm p}) - \delta_{\beta\gamma}\, \omega_{\alpha}^{(1)}({\bm k})\right)  = 0 
\label{eq:3.8a}
\ee 
with
\be
L^{(1)}_{\beta\gamma}({\bm k},{\bm p}) = \langle a_{\beta}({\bm p})\vert L_{\bm k}^{(1)}({\bm p})\vert a_{\gamma}({\bm p})\rangle A_{\gamma}^2
\label{eq:3.8b}
\ee
\ese
the elements of a matrix $L^{(1)}$. Due to the angular integration in the elements of $L^{(1)}$ the $5\times 5$ system decouples into
two scalar equations for the transverse velocity, or shear, modes and a $3\times 3$ system for
the density, the longitudinal velocity, and the energy or heat mode. Furthermore, the two shear
eigenvalues are the same by symmetry, $\omega_{3} = \omega_{4} = \omega_{\perp}$, and the angular integration makes
them vanish at this order,
\be
\omega_{\perp}^{(1)} = 0\ .
\label{eq:3.9}
\ee
One of the eigenvalues of the $3\times 3$ submatrix $L^{(1)}_L$ for the longitudinal modes ($\alpha, \beta = 1,2,5$)
also vanishes at this order,
\be
\omega_5^{(1)} = 0\ .
\label{eq:3.10}
\ee
This eigenvalue corresponds to the heat mode, see below. The other two, which correspond to first sound, have the form 
\be
\omega_{1,2}^{(1)} = \pm i c_1 k \ ,
\label{eq:3.11}
\ee
In order to determine the speed of first sound, $c_1$, we need the matrix $L_L^{(1)}$ explicitly. Evaluating the
matrix elements in Eq.~(\ref{eq:3.8b}) we find
\bse
\label{eqs:3.12}
\bea
L^{(1)}_L &=& \frac{ik}{3}\begin{pmatrix}      0                          & \frac{\langle {\bm v}_p\vert {\bm p}\rangle}{nm}   &    0                \\
            \frac{\langle {\bm v}_p\vert {\bm p}\rangle}{(\partial n/\partial\mu)_{T,V}}   &     0    &    \frac{\langle {\bm v}_p \vert {\bm p}\,a_5({\bm p})\rangle}{c_VT}  \\
                                            0                                         &  \frac{\langle {\bm v}_p \vert {\bm p}\, a_5({\bm p})\rangle}{nm} &    0                
                               \end{pmatrix}
\nonumber\\                               
\label{eq:3.12a}\\
\nonumber\\
&=& ik \begin{pmatrix}     0     & 1/m    &     0     \\
                                       \left(\frac{\partial p}{\partial n}\right)_{T,V}  &          0                    &   \frac{1}{c_V}\left(\frac{\partial p}{\partial T}\right)_{N,V}   \\
                                          0      & \frac{T}{nm} \left(\frac{\partial p}{\partial T}\right)_{N,V} &    0
             \end{pmatrix}\ ;
\nonumber\\             
\label{eq:3.12b}
\eea
\ese
%where $s$ is the equilibrium entropy density of the Fermi liquid, Eq.~(\ref{eq:A.11}). 
see Appendix~\ref{app:A.3} for the determination of the matrix elements in Eq.~(\ref{eq:3.12b}). 
We note that Eq.~(\ref{eq:3.12b}) represents the linearized Euler equations for an inviscid 
fluid.\cite{Landau_Lifshitz_VI_1959, Chaikin_Lubensky_1995} This illustrates that the theory so far
is consistent with general hydrodynamics; we will see below that this remains true if one takes into
account dissipation. Using Eq.~(\ref{eq:3.12b}) %and (\ref{eqs:2.21}) 
in (\ref{eq:3.8a}) we find
\bea
c_1^2 &=& \frac{1}{mn\chi_T} \left[1 + \frac{T \chi_T}{c_V}\left(\frac{\partial p}{\partial T}\right)_{N,V}^2 \right]
\nonumber\\
          &=& 1/mn\chi_S\ ,
\label{eq:3.13}
\eea
where $\chi_T = (-1/V)(\partial V/\partial p)_{T,N}$ is the isothermal compressibility, and $\chi_S = \chi_T c_V/c_p$ is the 
adiabatic one. We have used Eqs.~(\ref{eq:A.20b}) and (\ref{eq:A.27b}) to cast the speed of sound in this form.
Note that it is the coupling to the heat mode that changes the isothermal compressibility into the adiabatic one.
This is the correct result for the speed of sound; it is identical with the expressions obtained in the theory of classical 
fluids,\cite{Forster_1975} and in a phenomenological treatment of a Fermi gas.\cite{Fetter_Walecka_1971} However, 
we stress that it is not obvious {\it a priori} that LFL theory is consistent with general hydrodynamics in complete generality, 
not just at low temperatures. At $T=0$ we recover the well-known result\cite{Landau_Lifshitz_IX_1991, Baym_Pethick_1991}
\be
c_1^2 = \frac{1}{3}\,(\vF^*)^2 (1+F_0)(1+F_1/3)\ .
\label{eq:3.14}
\ee

\subsubsection{Hydrodynamic modes}
\label{subsubsec:III.A.2}

From Eqs.~(\ref{eq:3.6}) and (\ref{eq:3.8a}) we see that the left eigenvectors of $L^{(1)}$ determine the eigenmodes $\psi_{\alpha}^{\text{L}(0)}$ that
correspond to the eigenvalues $\omega_{\alpha}^{(1)}$. The shear modes decouple from the longitudinal modes
and from each other. The remaining three eigenmodes are determined by the left eigenvectors of the matrix $L_L^{(1)}$. Note that the
matrix is not symmetric, so we need to distinguish between left and right eigenvectors. 

\paragraph{Shear modes}
\label{par:III.A.2.a}

Since the shear modes decouple there is no difference between left and right, and we have
\bse
\label{eqs:3.15}
\be
\psi_{3,4}^{\text{L}(0)}({\bm p}) = \psi_{3,4}^{\text{R}(0)}({\bm p}) \equiv  \psi_{3,4}^{(0)}({\bm p}) = a_{3,4}({\bm p}) 
%\nonumber\\
= \hat{\bm k}_{\perp}^{1,2} \cdot {\bm p}\ .
\label{eq:3.15a}
\ee
The normalization is
\be
\left\langle\psi_{3,4}^{(0)}({\bm p})\vert\psi_{3,4}^{(0)}({\bm p})\right\rangle = A_2^{-2} = nm^*\ .
\label{eq:3.15b}
\ee
\ese

\paragraph{Heat mode}
\label{par:III.A.2.b}

The left and right eigenvectors associated with the zero eigenvalue $\omega_1^{(5)} = 0$ are not the same. For the left eigenvector we find
\bea
\psi_5^{\text{L}(0)}({\bm p}) &=& a_5({\bm p}) - \frac{T}{n} \left(\frac{\partial p}{\partial T}\right)_{\!\! N,V} a_1({\bm p})
\nonumber\\
&=&\epsilon_p  - (Ts/n + \mu)
\nonumber\\
&=&\epsilon_p - \frac{e+p}{n}\ ,
\label{eq:3.16}
\eea
where $s = S/V$ is the entropy density. 
Here we have used Eqs.~(\ref{eq:A.7}) and (\ref{eq:A.24}) to go from the first line to the second one; see also Eq.~(\ref{eq:2.19g}). The third line follows from the
general identity $Ts = e + p - n\mu$, with $e$ the energy density. We recognize $(e+p)/n$ as the enthalpy per particle. 

Equation~(\ref{eq:3.16}) shows that the thermodynamic quantity that represents the heat mode is 
\be
q({\bm x},t) = e({\bm x},t) - \frac{e+p}{n}\,n({\bm x},t)\ .
\label{eq:3.17}
\ee
which we recognize as the same quantity that represents the heat mode in a classical fluid.\cite{Forster_1975, Chaikin_Lubensky_1995}
This is as it should be, since the hydrodynamic arguments that lead to this mode are completely general. One can discern the
physical meaning of $q({\bm x},t)$ by considering the fluctuation of the entropy per particle. From the expression (\ref{eq:A.11}) for
the entropy density of a Fermi liquid we obtain
\bea
\delta(s/n)({\bm x},t) &=& \frac{1}{n}\,\delta s({\bm x},t) - \frac{s}{n^2}\,\delta n({\bm x},t) 
\nonumber\\
&=& \frac{1}{Tn}\,\frac{1}{V}\sum_{\bm p}\left(\epsilon_p-\mu - sT/n\right) \delta f({\bm p},{\bm x},t)
\nonumber\\
&=& \frac{1}{Tn}\,\frac{1}{V}\sum_{\bm p}\left(\epsilon_p - (e+p)/n\right) \delta f({\bm p},{\bm x},t)\ ,
\nonumber\\
\label{eq:3.18}
\eea
We thus have
\be
 \langle\psi^{\text{L}(0)}_5({\bm p})\vert\phi({\bm p},{\bm x},t)\rangle = T n\, \delta(s/n)({\bm x},t)\ ,
\label{eq:3.19}
\ee
which identifies the heat mode as the fluctuation of the entropy per particle.

To determine the right eigenvector we consider the right eigenproblem that corresponds to Eqs.~(\ref{eqs:3.8}). The relevant matrix is
\be
{\tilde L}^{(1)}_{\beta\gamma}({\bm k},{\bm p}) = A_{\beta}^2 \langle a_{\beta}({\bm p})\vert L_{\bm k}^{(1)}({\bm p})\vert a_{\gamma}({\bm p})\rangle \label{eq:3.20}
\ee
and the longitudinal submatrix is
\be
{\tilde L}^{(1)}_L = ik \begin{pmatrix}     0     &  \!\!\!\!\!  \frac{A_1^2}{A_2^2} \,\frac{1}{m} &    \!\!\!\!\!  0     \\
                         \frac{A_2^2}{A_1^2} \left(\frac{\partial p}{\partial n}\right)_{T,V}  &       \!\!\!\!\!      0        &  \!\!\!\!\!  \frac{A_2^2}{A_5^2}\frac{1}{c_V}\left(\frac{\partial p}{\partial T}\right)_{N,V}  \\
                                          0      &    \!\!\!\!  \frac{A_5^2}{A_2^2}\,\frac{T}{nm} \left(\frac{\partial p}{\partial T}\right)_{N,V} &   \!\!\!\!\!  0
             \end{pmatrix}
\label{eq:3.21}
\ee
The right eigenvector that corresponds to the eigenvalue $\omega_5^{(1)}=0$ is
\bse
\label{eqs:3.22}
\bea
\psi_5^{\text{R}(0)}({\bm p}) &=&  a_5({\bm p}) - \frac{T}{\langle 1\vert 1 \rangle}\,\frac{(\partial p/\partial T)_{N,V}}{(\partial p/\partial n)_{T,V}}\,a_1({\bm p})
\nonumber\\
&=& a_5({\bm p}) + \frac{T}{\langle 1\vert 1 \rangle}\, \left(\frac{\partial n}{\partial T}\right)_{p,V}\,a_1({\bm p})
\label{eq:3.22a}\\
&=&  a_5({\bm p}) - \frac{T}{n} \left(\frac{\partial p}{\partial T}\right)_{\!\! N,V} \frac{1}{1+F_0}\,a_1({\bm p})\ .\qquad\quad
\label{eq:3.22b}
\eea
\ese
Here we have used Eqs.~(\ref{eqs:A.20}) to arrive at Eq.~(\ref{eq:3.22b}). 
For the normalization of the heat mode this yields
\be
\langle \psi_5^{\text{L}(0)}({\bm p})\vert \psi_5^{\text{R}(0)}({\bm p})\rangle = T c_p
\label{eq:3.23}
\ee
where we have used Eqs.~(\ref{eq:A.20b}) and (\ref{eq:A.27b}).

\paragraph{Sound modes}
\label{par:III.A.2.c}

For the left eigenvectors associated with the eigenvalues $\omega^{(1)}_{1,2}$ we find, from Eq.~(\ref{eq:3.12b}),
%\begin{widetext}
\bse
\label{eqs:3.24}
\bea
\psi^{\text{L}(0)}_{1,2}({\bm p}) &=&  \pm c_1 a_2({\bm p}) + \frac{1}{c_V} \left(\frac{\partial p}{\partial T}\right)_{\!\! N,V} a_5({\bm p})
\nonumber\\
     && \hskip 45pt + \left(\frac{\partial p}{\partial n}\right)_{\! T,V} a_1({\bm p})
\label{eq:3.24a}\\
&&\hskip -50pt  =  \pm c_1 a_2({\bm p}) + \frac{1}{c_V} \left(\frac{\partial p}{\partial T}\right)_{\!\! N,V} \psi_5^{{\text L}(0)}({\bm p}) + m c_1^2 a_1({\bm p})\ . 
\nonumber\\
\label{eq:3.24b}
\eea
\ese
%\end{widetext}
Here we have used Eqs.~(\ref{eq:3.16}) and (\ref{eq:3.14}) to arrive at the second line. Combining Eqs.~(\ref{eq:2.19f}), (\ref{eq:2.19h}), and
(\ref{eq:3.24a}) we see that the sound modes are linear combinations of longitudinal velocity fluctuations, Eq.~(\ref{eq:2.19b}), and pressure 
fluctuations $\delta p$:
\be
\langle\psi^{\text{L}(0)}_{1,2}({\bm p})\vert\phi({\bm p},{\bm x},t)\rangle = \pm c_1 m\, \delta u_{\text{L}}({\bm x},t) + \delta p({\bm x},t)\ . 
\label{eq:3.25}
\ee
The corresponding right eigenvector is obtained from Eq.~(\ref{eq:3.21}) as
\bse
\label{eqs:3.26}
\bea
\psi^{\text{R}(0)}_{1,2}({\bm p}) = \pm c_1 a_2({\bm p}) 
     &+& \frac{m^*}{c_V m} \left(\frac{\partial p}{\partial T}\right)_{N,V} a_5({\bm p}) 
\nonumber\\     
     &+& \frac{n m^*}{m} \frac{1}{\langle 1\vert 1\rangle}\,a_1({\bm p})\ ,
     \label{3.26a}
\eea
and for the normalization we obtain
\be
\langle \psi^{\text{L}(0)}_{1,2}({\bm p})\vert\psi^{\text{R}(0)}_{1,2}({\bm p})\rangle = 2 n m^* c_1^2
\label{eq:3.26b}
\ee
\ese
where we have used Eqs.~(\ref{eq:A.20b}) and (\ref{eq:3.13}).

All of these results are consistent with the corresponding ones for a classical 
fluid.\cite{Ernst_Hauge_van_Leeuwen_1976a, Dorfman_vanBeijeren_Kirkpatrick_2021}
Note that the hydrodynamic modes are all orthogonal to one another, as they must be:
\be
\big\langle\psi_{\alpha}^{\text{L}(0)}({\bm p})\big\vert\psi_{\beta}^{\text{R}(0)}({\bm p})\big\rangle = 
     \delta_{\alpha\beta}\, \big\langle\psi_{\alpha}^{\text{L}(0)}({\bm p})\big\vert\psi_{\alpha}^{\text{R}(0)}({\bm p})\big\rangle\ .
\label{eq:3.27}
\ee
With the zeroth order eigenmodes $\psi^{(0)}$ in place, Eq.~(\ref{eq:3.7}) now constitutes a well-defined integral equation
that determines the modes at $O(k)$. The solution of this equation is not unique since one can add an
arbitrary linear combination of the $a_{\alpha}$ to $\psi_{\alpha}^{L(1)}$ and still satisfy the equation. 
Uniqueness is achieved by the requirement that $\psi_{\alpha}^{L(1)}$ be an element of the orthogonal
space ${\cal L}_{\perp}$. We can formally write this unique solution as
\bse
\label{eqs:3.28}
\be
\langle\psi_{\alpha}^{\text{L}(1)}({\bm p})\vert = 
\langle \psi_{\alpha}^{\text{L}(0)}({\bm p})\vert\, \left[ L^{(1)}_{\bm k}({\bm p}) - \omega_{\alpha}^{(1)}({\bm k})\right]
\Lambda^{-1}({\bm p})\,{\cal P}_{\perp}\ . \qquad\quad
\label{eq:3.28a}
\ee
Here the projection operator ${\cal P}_{\perp}$ projects onto the orthogonal space ${\cal L}_{\perp}$, and the inverse collision operator
formally exists since it acts on a vector in ${\cal L}_{\perp}$. The formal expression (\ref{eq:3.28a}) should be interpreted as the solution
of the integral Eq.~(\ref{eq:3.7}) made unique by the orthogonality requirement that is enforced by ${\cal P}_{\perp}$.  
The corresponding right eigenvector at this order is
\be
\vert\psi_{\alpha}^{\text{R}(1)}({\bm p})\rangle = {\cal P}_{\perp} \Lambda^{-1}({\bm p}) \left[L_{\bm k}^{(1)}({\bm p} - \omega_{\alpha}^{(1)}({\bm k})\right] \vert \psi_{\alpha}^{\text{R}(0)}({\bm p})\rangle\ .
\label{eq:3.28b}
\ee
\ese

\begin{widetext}
\subsubsection{Shear diffusion, heat diffusion, and sound attenuation coefficients}
\label{subsubsec:III.A.3}

We now consider the eigenproblem, Eq.~(\ref{eq:3.1}), at second order in $k$. At this order we have
\be
- \langle\psi_{\alpha}^{\text{L}(2)}({\bm p})\vert \, \Lambda({\bm p}) + \langle\psi_{\alpha}^{\text{L}(1)}({\bm p})\vert \left(L_{\bm k}^{(1)}({\bm p}) 
     - \omega_{\alpha}^{(1)}({\bm k})\right) 
%\nonumber\\
=  \langle\psi_{\alpha}^{\text{L}(0)}({\bm p})\vert\,\omega_{\alpha}^{(2)}({\bm k},{\bm p})\ .
\label{eq:3.29}
\ee
If we multiply with $\vert\psi_{\alpha}^{\text{R}(0)}\rangle$ from the right and use Eq.~(\ref{eq:3.28a}) this becomes
%\begin{widetext}
\be
\omega_{\alpha}^{(2)}({\bm k}) = \frac{1}{\big\langle\psi_{\alpha}^{\text{L}(0)}({\bm p})\big\vert\psi_{\alpha}^{\text{R}(0)}({\bm p})\big\rangle}\,
  \langle\psi_{\alpha}^{\text{L}(0)}({\bm p})\vert \left(L_{\bm k}^{(1)}({\bm p}) - \omega_{\alpha}^{(1)}({\bm k})\right) 
\Lambda^{-1}({\bm p})  \left(L_{\bm k}^{(1)}({\bm p}) - \omega_{\alpha}^{(1)}({\bm k})\right) \vert\psi_{\alpha}^{\text{R}(0)}({\bm p})\rangle\ .
\label{eq:3.30}
\ee
\end{widetext}

\paragraph{Shear diffusion}
\label{par:III.A.3.a}

We again first consider the shear modes, $\alpha = 3,4$. In this case $\omega_{3,4}^{(1)} \equiv \omega_{\perp}^{(1)} = 0$, and we 
obtain for the shear eigenvalue at second order
\bse
\label{eqs:3.31}
\bea
\omega_{\perp}^{(2)} \!\!&=&\!\! \frac{1}{nm^*}\,  \langle\psi_{\perp}^{(0)}({\bm p})\vert L_{\bm k}^{(1)}({\bm p}) 
\Lambda^{-1}({\bm p}) L_{\bm k}^{(1)}({\bm p}) \vert\psi_{\perp}^{(0)}({\bm p})\rangle
\nonumber\\
\label{eq:3.31a}\\
&=& \nu k^2\ .
\label{eq:3.31b}
\eea
\ese
Here $\psi_{\perp}^{(0)}$ stands for either $\psi_3^{(0)}$ or $\psi_4^{(0)}$, and $\nu$ is the kinematic viscosity, which 
is given by the matrix element in Eq.~(\ref{eq:3.31a}). It
is related to the shear viscosity $\eta$ via the physical mass density:
\be
\nu = \eta/nm \ .
\label{eq:3.32}
\ee
This is an Einstein relation that relates the kinematic viscosity $\nu$, which is the shear diffusion coefficient,
and the transport coefficient $\eta$ via the static transverse momentum susceptibility, which is given by 
$nm$.

Using the $L_{\bm k}^{(1,2)}$ part of the kinetic operator in the form of Eq.~(\ref{eq:3.4d}) we find
\bse
\label{eqs:3.33}
\bea
L_{\bm k}^{(1)}({\bm p}) \big\vert\psi_{\perp}^{(0)}({\bm p})\big\rangle &=& (1+F_1/3) \big\vert i({\bm k}\cdot{\bm v}_p)(\hat{\bm k}_{\perp}\cdot{\bm p})\big\rangle\ ,
\nonumber\\
\label{eq:3.33a}\\
\big\langle\psi_{\perp}^{(0)}({\bm p})\big\vert L_{\bm k}^{(1)}({\bm p}) &=& \big\langle i({\bm k}\cdot{\bm v}_p)(\hat{\bm k}_{\perp}\cdot{\bm p})\big\vert\ ,
\label{eq:3.33b}
\eea
\ese
where $\hat{\bm k}_{\perp}$ stands for either $\hat{\bm k}_{\perp}^1$ or $\hat{\bm k}_{\perp}^2$. Note that both of these
vectors are orthogonal to all five vectors $a_{\alpha}({\bm p})$ that span the zero-eigenvector space ${\cal L}_0$, as they must be, so the matrix
element in Eq.~(\ref{eq:3.31a}) exists. We thus have
\be
\eta = -\left\langle(\hat{\bm k}_{\perp}\cdot{\bm p})(\hat{\bm k}\cdot{\bm v}_p) \big\vert \Lambda^{-1}({\bm p}) \big\vert 
      (\hat{\bm k}\cdot{\bm v}_p)(\hat{\bm k}_{\perp}\cdot{\bm p})\right\rangle\ .
\label{eq:3.34}
\ee
That is,
the shear viscosity is given by a transverse stress correlation, as expected from classical kinetic theory
at the level of the Boltzmann equation.\cite{Dorfman_vanBeijeren_Kirkpatrick_2021} 

If we replace the collision operator by the BKG model collision operator from Appendix~\ref{app:C}, then $\Lambda^{-1}({\bm p})$ in 
Eq.~(\ref{eq:3.34}) effectively becomes $-\tau$, with $\tau$ the collision time. At $T=0$ we then
recover the expression for the kinematic viscosity and the shear viscosity familiar from LFL theory:\cite{Baym_Pethick_1991}
\bse
\label{eqs:3.35}
\bea
\nu &=& \frac{1}{5}\,(\vF^*)^2\tau(1+F_1/3)\ ,
\label{eq:3.35a}\\
\eta &=&  \frac{nm}{5}\,(\vF^*)^2\tau (1+F_1/3)\ .
\label{eq:3.35b}
\eea
\ese

\paragraph{Heat diffusion}
\label{par:III.A.3.b}

Equation~(\ref{eq:3.30}) for $\alpha = 5$ yields
\bse
\label{eqs:3.36}
\bea
\omega_{5}^{(2)} \!\!\!&=&\!\! \frac{1}{T c_p}  \langle\psi_{5}^{\text{L}(0)}({\bm p})\vert L_{\bm k}^{(1)}({\bm p}) 
\Lambda^{-1}({\bm p}) L_{\bm k}^{(1)}({\bm p}) \vert\psi_{5}^{\text{R}(0)}({\bm p})\rangle
\nonumber\\
\label{eq:3.36a}\\
&=& D_{T} k^2
\label{eq:3.36b}
\eea
\ese
where $D_T$ is the heat diffusion coefficient. It is related to the thermal or heat conductivity $\kappa$ via
\be
D_T = \kappa/c_p \ .
\label{eq:3.37}
\ee

Using Eq.~(\ref{eq:3.4d}) to calculate the vectors in the matrix element in Eq.~(\ref{eq:3.36b}) we find
\bse
\label{eqs:3.38}
\bea 
\big\langle\psi_{5}^{\text{L}(0)}({\bm p})\big\vert L_{\bm k}^{(1)}({\bm p}) &=& 
     ik\, \big\langle\psi_{5}^{\text{L}(0)}({\bm p})(\hat{\bm k}\cdot{\bm v}_p)\big\vert\ , \qquad\quad
\label{eq:3.38a}\\
 L_{\bm k}^{(1)}({\bm p}) \big\vert\psi_{5}^{\text{R}(0)}({\bm p})\big\rangle &=& 
    ik\, \big\vert (\hat{\bm k}\cdot{\bm v}_p) \psi_{5}^{\text{L}(0)}({\bm p})\big\rangle\ .
 \label{eq:3.38b}
 \eea
 \ese
Note that the kinetic operator $L_{\bm k}^{(1)}$ turns the right eigenfunction, Eq.~(\ref{eqs:3.22}) into a current related to the left 
eigenfunction, given by Eq.~(\ref{eq:3.16}). We have used Eqs.~(\ref{eq:A.4}) and (\ref{eqs:A.20}) to arrive at this result.
For the thermal conductivity this yields
\be
\kappa = \frac{-1}{T} \big\langle\psi_{5}^{\text{L}(0)}({\bm p})(\hat{\bm k}\cdot{\bm v}_p)\big\vert \Lambda^{-1}({\bm p})
     \big\vert (\hat{\bm k}\cdot{\bm v}_p) \psi_{5}^{\text{L}(0)}({\bm p})\big\rangle\ .
\label{eq:3.39}     
\ee
Again, this is consistent with the corresponding result in classical kinetic theory,\cite{Dorfman_vanBeijeren_Kirkpatrick_2021} 
and Eq.~(\ref{eq:3.37}) is an Einstein relation that relates the heat diffusion coefficient $D_T$ to the transport
coefficient $T\kappa$ via the heat susceptibility $T c_p$.

If we replace $\Lambda({\bm p})$ by the BGK model collision operator from Appendix~\ref{app:C} and evaluate 
Eq.~(\ref{eq:3.39}) to lowest order in the temperature, we find
\bse
\label{eqs:3.40}
\bea
\kappa &=& c_p\,\frac{1}{3} (\vF^*)^2\tau\ ,
\label{eq:3.40a}\\
D_T &=& \frac{1}{3} (\vF^*)^2\tau\ ,
\label{eq:3.40b}
\eea
\ese
which is the result familiar from LFL theory.\cite{Baym_Pethick_1991} Note that at this order there is no difference between
$c_p$ and $c_V$. 

\paragraph{Sound attenuation}
\label{par:III.A.3.b}

We finally consider Eq.~(\ref{eq:3.30}) for $\alpha = 1,2$. We have
\begin{widetext}
\bse
\label{eqs:3.41}
\bea
\omega_{1,2}^{(2)} &=& \frac{1}{2nm^* c_1^2} \big\langle \psi_{1,2}^{\text{L}(0)}({\bm p})\big\vert \left( L_{\bm k}^{(1)}({\bm p}) \mp i c_1 k\right)
   \Lambda^{-1}({\bm p})  \left( L_{\bm k}^{(1)}({\bm p}) \mp i c_1 k\right) \big\vert \psi_{1,2}^{\text{R}(0)}({\bm p})\big\rangle
\label{eq:3.41a}\\
&=& \frac{1}{2}\,\Gamma\,k^2
\label{eq:3.41b}
\eea
\ese
with $\Gamma$ the sound attenuation coefficient.

Using various identities from Appendix~\ref{app:A}, as well as Eq.~(\ref{eq:2.10a}), we can write the relevant left vector in the form
\bse
\label{eqs:3.42}
\be
\big\langle \psi_{1,2}^{\text{L}(0)}({\bm p})\big\vert \left( L_{\bm k}^{(1)}({\bm p}) \mp i c_1 k\right) = ik \left[
     \pm c_1\, {\hat k}_i{\hat k}_j \big\langle\sigma^{ij}({\bm p}) \big\vert 
     +  \frac{1}{c_V}\left(\frac{\partial p}{\partial T}\right)_{N,V} \big\langle\psi_5^{\text{L}(0)}({\bm p}) (\hat{\bm k}\cdot{\bm v}_p) \big\vert
     + mc_1^2  \big\langle\hat{\bm k}\cdot({\bm v}_p - {\bm p}/m^*) \big\vert      \right] \ ,
\label{eq:3.42a}
\ee
where we have used the sound mode in the form (\ref{eq:3.24b}).
The second term in Eq.~(\ref{eq:3.42a}) we recognize as proportional to the heat
current density from Eqs.~(\ref{eqs:3.38}) and Appendix~\ref{app:B.2}. For the corresponding right vector we find
\be
\left( L_{\bm k}^{(1)}({\bm p}) \mp i c_1 k\right) \big\vert \psi_{1,2}^{\text{R}(0)}({\bm p})\big\rangle = i k\,\frac{m^*}{m} \left[ 
     \pm c_1\,{\hat k}_i {\hat k}_j  \big\vert\sigma^{ij}({\bm p})\big\rangle 
     + \frac{1}{c_V}\left(\frac{\partial p}{\partial T}\right)_{N,V} \big\lvert\psi_5^{\text{L}(0)}({\bm p}) (\hat{\bm k}\cdot{\bm v}_p) \big\rangle
          + mc_1^2  \big\vert\hat{\bm k}\cdot({\bm v}_p - {\bm p}/m^*) \big\rangle     \right]\ .
\label{eq:3.42b}
\ee
\ese
\end{widetext}
The last term on the right-hand side of both Eq.~(\ref{eq:3.42a}) and (\ref{eq:3.42b}) is the component of the
longitudinal QP velocity in the orthogonal space ${\cal L}_{\perp}$. It thus vanishes for the model interaction given
by Eq.~(\ref{eq:3.4d}), and we drop it. The stress tensor $\sigma^{ij}$ is given by
\bse
\label{eqs:3.43}
\be
\sigma^{ij}({\bm p}) = \sigma_1^{ij}({\bm p}) - \delta^{ij}\,\sigma_2({\bm p})
\label{eq:3.43a}
\ee
where
\be
\sigma_1^{ij}({\bm p}) = p^i v_p^j - \delta^{ij} \frac{1}{3}\,{\bm p}\cdot{\bm v}_p\ ,
\label{eq:3.43b}
\ee
and
\bea
\sigma_2({\bm p}) &=& \frac{1}{c_V}\left(\frac{\partial p}{\partial T}\right)_{N,V} \epsilon_p - \frac{1}{3}\,{\bm p}\cdot{\bm v}_p
\nonumber\\
&& + \frac{n}{\langle 1\vert 1\rangle} - \frac{1}{c_V}\left(\frac{\partial p}{\partial T}\right)_{N,V} \langle\epsilon_p\rangle_w\ .
\label{eq:3.43c}
\eea
\ese 
Note that $\sigma_1$ and $\sigma_2$ are separately orthogonal to the zero-eigenvector space. To calculate the eigenfrequencies,
Eqs.~(\ref{eqs:3.41}), we use the fact that the collision operator is isotropic in momentum space. As a result, the contributions
from $\sigma_1$, $\sigma_2$, and the heat current do not mix, and the contribution from $\sigma_1$ can be related to the
shear viscosity $\eta$, Eq.~(\ref{eq:3.34}). After a calculation that makes extensive use of the identities in Appendix~\ref{app:A}, 
in particular Eqs.~(\ref{eqs:A.20}) and (\ref{eqs:A.27}), we obtain
\bse
\label{eqs:3.44}
\be
\Gamma = \frac{1}{nm} \left(\frac{4}{3}\,\eta + \zeta \right) + D_T \left(\frac{c_p}{c_V} - 1 \right)\ .
\label{eq:3.44a}
\ee
Here $D_T$ is the heat diffusion coefficient, Eqs.~(\ref{eq:3.37}, \ref{eq:3.39}), $\eta$ is the shear viscosity, Eq.~(\ref{eq:3.34}), and 
\be
\zeta = - \langle \sigma_2({\bm p})\vert\Lambda^{-1}({\bm p})\vert \sigma_2({\bm p})\rangle
\label{eq:3.44b}
\ee
\ese
is a contribution to the bulk viscosity.\cite{bulk_viscosity_footnote} The latter vanishes in a Fermi gas, where $p = 2e/3$ and hence 
$\sigma_2({\bm p})=0$. It also vanishes in the model where the QP interaction is given by Eq.~(\ref{eq:3.4d}).
This can be seen by realizing that $-3\vert\sigma_2({\bm p})\rangle = {\cal P}_{\perp} \vert {\bm p}\cdot{\bm v}_p\rangle$ is
the projection of ${\bm p}\cdot{\bm v}_p$ onto the orthogonal space ${\cal L}_{\perp}$. But from Eq.~(\ref{eq:2.10b}) it follows
that ${\bm p}\cdot{\bm v}_p \in {\cal L}_0$, and hence $\sigma_2({\bm p}) = 0$.
However, in the case of a more general QP interaction it 
will be nonzero and of $O(T^4)$ relative to the shear viscosity, see the remarks in Ref.~\onlinecite{v_p_footnote}. 
These properties of $\zeta$ are consistent  with the results of
Ref.~\onlinecite{Sykes_Brooker_1970}, and the expression for the sound attenuation, Eq.~(\ref{eq:3.44a}), is the same as the 
one for a classical fluid.\cite{Forster_1975}

The complete result for the eigenvalues $\omega_{1,2}$ that describe first sound is
\be
\omega_{1,2}({\bm k}) = \pm i c_1 k + \frac{1}{2}\,\Gamma\,k^2 + O(k^3)\ ,
\label{eq:3.45}
\ee
with $c_1$ the speed of sound from Eq.~(\ref{eq:3.13}), and $\Gamma$ the sound attenuation coefficient from Eq.~(\ref{eq:3.44a}).

\smallskip
\subsection{Long-ranged case}
\label{subsec:III.B}

As mentioned in the context of Eq.~(\ref{eq:2.11}), in a charged Fermi liquid the interaction parameter $F_0$ acquires an
additional contribution that is due to the Coulomb interaction and diverges for small wave numbers as $1/k^2$. The main
effect of this term is to turn the soft sound mode into the massive (in $d=3$) plasmon mode. It also affects the heat diffusion 
coefficient, but does not change the diffusive nature of the heat transport. 

We define the plasma frequency $\omega_p$
\bse
\label{eqs:3.46}
\be
\omega_p^2 = 4\pi n e^2/m\ ,
\label{eq:3.46a}
\ee
and the Thomas-Fermi screening wave number $p_{\text{TF}}$,
\be
p_{\text{TF}}^2 = n m \chi_T\, \omega_p^2\ .
\label{eq:3.46b}
\ee
\ese
Here $\chi_T = (\partial n/\partial\mu)_{N,V}/n^2 = N_0/n^2(1+F_0)$ is the isothermal compressibility in the SR case. 
The isothermal compressibility of the Coulomb system takes the form
\be
\chi_T^{\text{LR}} = \chi_T\,\frac{k^2}{k^2 + p_{\text{TF}}^2}\ .
\label{eq:3.47}
\ee
Note that $\chi_T^{\text{LR}}$ is wave-number dependent and vanishes at $k=0$, i.e. the system is
incompressible with respect to uniform deformations.

\subsubsection{Hydrodynamic modes}
\label{subsubsec:III.B.1}

The linearized kinetic operator is now given by
\be
L_{\bm k}({\bm p}) = L_{\bm k}^{(-1)}({\bm p}) - \Lambda({\bm p}) + L_{\bm k}^{(1)}({\bm p})\ ,
\label{eq:3.48}
\ee
with $L_{\bm k}^{(1)}$ from Eqs.~(\ref{eqs:3.4}) and
\be
L_{\bm k}^{(-1)}({\bm p}) = \frac{4\pi e^2}{k^2}\,\big\vert i{\bm k}\cdot{\bm v}_p\big\rangle \big\langle 1\big\vert
\label{eq:3.49}
\ee
Accordingly, the left eigenproblem from Eqs.~(\ref{eqs:3.8}) needs to be augmented by a matrix
\be
L^{(-1)}_{\beta\gamma}({\bm k},{\bm p}) = \langle a_{\beta}({\bm p})\vert L_{\bm k}^{(-1)}({\bm p})\vert a_{\gamma}({\bm p})\rangle A_{\gamma}^2
\label{eq:3.50}
\ee
whose only nonzero element is $L^{(-1)}_{21} ({\bm k},{\bm p}) = i\,\omega_p^2\, m/k$. As a result, the matrix $L_L$ for the longitudinal
modes acquires a contribution of $O(1/k)$, and instead of Eq.~(\ref{eq:3.12b}) we have
\begin{widetext}
\be
L_L^{(-1)} + L_L^{(1)} =  ik \begin{pmatrix}     0     & 1/m    &   0     \\
                                       \left(\frac{\partial p}{\partial n}\right)_{T,V} + \frac{m\omega_p^2}{k^2}  &   0   & \frac{1}{c_V}\left(\frac{\partial p}{\partial T}\right)_{N,V}   \\
                                          0      & \frac{T}{nm} \left(\frac{\partial p}{\partial T}\right)_{N,V} & 0
             \end{pmatrix}
\label{eq:3.51}
\ee   
\end{widetext}
This again represents the linearized Euler equations, albeit in the presence of a long-ranged interaction.        
The eigenvalue $\omega_5$, which corresponds to the heat mode, is still zero at this level. However, the eigenvalues
$\omega_{1,2}$ are now given by 
\bse
\label{eqs:3.52}
\be
\omega_{1,2}^2 = -\left(\omega_p^2 + c_1^2 k^2\right)
\label{eq:3.52a}
\ee
with $c_1$ the speed of first sound in the SR case, Eq.~(\ref{eq:3.13}). They thus have a contribution at $O(k^0)$,
\be
\omega_{1,2}^{(0)} = \pm i\omega_p\ ,
\label{eq:3.52b}
\ee
and one at $O(k^2)$,
\be
\omega_{1,2}^{(2)} = \pm i\,k^2\,c_1^2/2\omega_p\ ,
\label{eq:3.52c}
\ee
\ese
as well as contributions at higher order. This reflects the fact that the soft first-sound modes have been
turned into massive plasmon modes by the long-ranged interaction.\cite{mode-coupling_footnote}

\paragraph{Shear modes}
\label{par:III.B.1a}

The shear modes are unaffected by the long-ranged interaction and are still given by Eqs.~(\ref{eqs:3.15}). 

\paragraph{Heat mode}
\label{par:III.B.1b}

Similarly, the heat mode, i.e., the left eigenvector associated with the eigenvalue $\omega_5=0$, is unaffected and still
given by Eq.~(\ref{eq:3.16}). However, the corresponding right eigenvector, which we obtain
from the modification of Eq.~(\ref{eq:3.21}) that corresponds to Eq.~(\ref{eq:3.51}), now has one
contribution at $O(k^0)$ and another at  $O(k^2)$, as can be seen from Eq.~(\ref{eq:3.22b}) in conjunction
with Eq.~(\ref{eq:2.11}). The components of the eigenvectors in the zero eigenvector space ${\cal L}_0$ thus are
\bse
\label{eqs:3.53}
\bea
\psi_5^{\text{L}(0)}({\bm p}) &=& a_5({\bm p}) - \frac{T}{n} \left(\frac{\partial p}{\partial T}\right)_{\!\! N,V} a_1({\bm p})\ ,
\label{eq:3.53a}\\
\psi_5^{\text{R}(0)}({\bm p}) &=& a_5({\bm p})\ ,
\label{eq:3.53b}\\
\psi_5^{\text{R}(2)}({\bm p}) &=& -k^2\,\frac{T(\partial p/\partial T)_{N,V}}{m\,\omega_p^2\,\langle 1\vert 1\rangle}\, a_1({\bm p}) .
\label{eq:3.53b}
\eea
\ese
Accordingly, the normalization of the heat mode to lowest order in $k$ is given by $c_V$ rather then $c_p$:
\be
\langle \psi_5^{\text{L}(0)}({\bm p})\vert \psi_5^{\text{R}(0)}({\bm p})\rangle = T c_V\ .
\label{eq:3.54}
\ee

\paragraph{Plasmon modes}
\label{par:III.B.1c}

The plasmon modes are given by the left eigenvectors of the matrix in Eq.~(\ref{eq:3.51}) for the eigenvalues $\omega_{1,2}$.
If we normalize the modes such that the generalization of Eq.~(\ref{eq:3.26b}) is finite at $k=0$, we have, up to $O(k^2)$,
\bse
\label{eqs:3.55}
\bea
\psi_{1,2}^{\text{L}(-1)}({\bm p}) &=& \frac{1}{k}\,m\,\omega_p^2\,a_1({\bm p})
\label{eq:3.55a}\\
\psi_{1,2}^{\text{L}(0)}({\bm p}) &=& \pm \omega_p\,a_2({\bm p}) 
\label{eq:3.55b}\\
\psi_{1,2}^{\text{L}(1)}({\bm p}) &=& k \left[ \frac{1}{n\chi_T}\, a_1({\bm p}) +
     \frac{1}{c_V}\left(\frac{\partial p}{\partial T}\right)_{N,V} a_5({\bm p})\right]
     \nonumber\\
\label{eq:3.55c}\\
\psi_{1,2}^{\text{L}(2)}({\bm p}) &=& \pm k^2\,\frac{c_1^2}{2\omega_p}\,a_2({\bm p})\ .
\label{eq:3.55d}
\eea
\ese
For the corresponding right eigenvector one finds
\bse
\label{eqs:3.56}
\bea
\psi_{1,2}^{\text{R}(0)}({\bm p}) &=& \pm\, \omega_p\,a_2({\bm p}) 
\label{eq:3.56a}\\
\psi_{1,2}^{\text{R}(1)}({\bm p}) &=& k\,\frac{m^*}{m} \left[\frac{n}{\langle 1\vert 1\rangle}\, a_1({\bm p}) 
             +  \frac{1}{c_V}\left(\frac{\partial p}{\partial T}\right)_{N,V} a_5({\bm p})\right]\ ,
             \nonumber\\
\label{eq:3.56b}\\
\psi_{1,2}^{\text{R}(2)}({\bm p}) &=& \pm k^2\,\frac{c_1^2}{2\omega_p}\,a_2({\bm p})\ .
\label{eq:3.56c}
\eea
\ese
The normalization now is
\be
\langle \psi_{1,2}^{\text{L}}({\bm p}) \vert \psi_{1,2}^{\text{R}}({\bm p}) \rangle = 2nm^*\omega_p^2 + O(k^2)\ .
\label{eq:3.57}
\ee

This completes the determination of the components of the hydrodynamic modes that are elements of ${\cal L}_0$.
In contrast to the SR case they do not all occur at the same order in the $k$-expansion.
Note that all of the results so far can be obtained from those for the SR case, 
Secs.~\ref{subsubsec:III.A.1} and \ref{subsubsec:III.A.2}, by using the substitution (\ref{eq:2.11}) for the Fermi-liquid 
parameter $F_0$.

\subsubsection{Transport coefficients, and plasmon damping and dispersion}
\label{subsubsec:III.B.2}

\paragraph{Shear diffusion}
\label{par:III.B.2a}

As we have seen, the shear modes are unaffected by the Coulomb interaction. Accordingly, the shear viscosity and the shear diffusion
coefficient are still given by Eqs.~(\ref{eq:3.34}) and (\ref{eq:3.32}), respectively. 

\smallskip
\paragraph{Heat diffusion}
\label{par:III.B.2b}

Consider the left eigenproblem, Eq.~(\ref{eq:3.1}), for $\alpha = 5$ with the kinetic operator given by Eq.~(\ref{eq:3.48}).
The parts of the eigenfunctions that are elements of the zero eigenvector space are given by Eqs.~(\ref{eqs:3.53}). 
Expanding in powers of $k$ as in Sec.~\ref{subsubsec:III.A.1} we have at lowest order
\bse
\label{eqs:3.58}
\be
\langle \psi_5^{\text{L}(0)}({\bm p})\vert\, L_{\bm k}^{(-1)}({\bm p}) = 0\ ,
\label{eq:3.58a}
\ee
which holds since the angular integration vanishes. At $O(k^0)$ we have
\be
- \langle \psi_5^{\text{L}(0)}({\bm p})\vert\,\Lambda({\bm p}) + \langle \psi_5^{\text{L}(1)}({\bm p})\vert\,L_{\bm k}^{(-1)}({\bm p}) = 0\ .
\label{eq:3.58b}
\ee
Here the first term on the left-hand side vanishes due to the conservation laws. The vector $\langle \psi_5^{\text{L}(1)}({\bm p})\vert$
is orthogonal to the zero eigenvector space, whereas the vector $\vert{\bm k}\cdot{\bm v}_p\rangle$ in the operator
$L_{\bm k}^{(-1)}$ is proportional to the zero eigenvector $\vert a_2({\bm p})\rangle$ due to Eq.~(\ref{eq:2.10a}). Hence this
condition is also fulfilled. Anticipating that the first nonzero contribution to the eigenvalue appears only at $O(k^2)$, at $O(k)$ we have
\bea
\langle \psi_5^{\text{L}(0)}({\bm p})\vert\,L_{\bm k}^{(1)}({\bm p}) &-& \langle \psi_5^{\text{L}(1)}({\bm p})\vert\,\Lambda({\bm p}) 
\nonumber\\
&+& \langle \psi_5^{\text{L}(2)}({\bm p})\vert\,L_{\bm k}^{(-1)}({\bm p}) = 0\ .\qquad\quad
\label{eq:3.58c}
\eea
\ese
The last term on the left-hand side again vanishes since $\psi_5^{\text{L}(2)}$ is orthogonal to the zero eigenvector space,
and we obtain a formal expression for $\psi_5^{\text{L}(1)}$ in analogy to Eq.~(\ref{eq:3.28a}):
\begin{widetext}
\bse
\label{eqs:3.59}
\be
\langle\psi_5^{\text{L}(1)}({\bm p})\vert = \langle \psi_5^{\text{L}(0)}({\bm p})\vert\,L_{\bm k}^{(1)}({\bm p})\,\Lambda^{-1}({\bm p})\,{\cal P}_{\perp}
%\nonumber\\
= ik \langle \psi_5^{\text{L}(0)}({\bm p}) (\hat{\bm k}\cdot{\bm v}_p)\vert\,\Lambda^{-1}({\bm p})\,{\cal P}_{\perp}\ .\qquad
\label{eq:3.59a}
\ee
Analogous arguments for the right eigenproblem yield
\be
\big\vert \psi_5^{\text{R}(1)}({\bm p})\big\rangle = {\cal P}_{\perp} \Lambda^{-1} \left(L_{\bm k}^{(-1)}({\bm p}) \big\vert  \psi_5^{\text{R}(2)}({\bm p})\big\rangle 
   +  L_{\bm k}^{(1)}({\bm p}) \big\vert  \psi_5^{\text{R}(0)}({\bm p})\big\rangle \right)
%\nonumber\\
= ik\,{\cal P}_{\perp} \Lambda^{-1}({\bm p}) \big\vert (\hat{\bm k}\cdot{\bm v}_p)\psi_5^{\text{L}(0)}({\bm p})\big\rangle\ .
\label{eq:3.59b}
\ee
\ese
Note that linear combination of vectors on the right-hand side of Eq.~(\ref{eq:3.59b}) produces the same result for 
$\psi_5^{\text{R}(1)}$ as in the SR case, see Eqs.~(\ref{eqs:3.38}).

In order to determine the eigenvalue at $O(k^2)$ we multiply Eq.~(\ref{eq:3.1}) from the right with $\vert\psi_5^{\text{R}}({\bm p})\rangle$
and expand order by order in powers of $k$. The terms at $O(k^{-1})$ and $O(k^0)$ vanish due to a combination of $L_{\bm k}^{(-1)}$
acting on the orthogonal space and the conservation laws. At $O(k)$ we find
\be
\omega_5^{(1)}({\bm k}) \langle\psi_5^{\text{L}(0)}({\bm p})\vert \psi_5^{\text{R}(0)}({\bm p})\rangle = \langle\psi_5^{\text{L}(0)}({\bm p}) \vert
L_{\bm k}^{(1)}({\bm p}) \vert \psi_5^{\text{R}(0)}({\bm p})\rangle\ ,
\label{eq:3.60}
\ee
which vanishes due to the angular integration, so 
\be
\omega_5^{(1)}({\bm k}) = 0
\label{eq:3.61}
\ee
as expected. At $O(k^2)$ we find
\bse
\label{eqs:3.62}
\be
\omega_5^{(2)}({\bm k}) \langle\psi_5^{\text{L}(0)}({\bm p})\vert \psi_5^{\text{R}(0)}({\bm p})\rangle =
     \langle\psi_5^{\text{L}(0)}({\bm p})\vert L_{\bm k}^{(1)}({\bm p}) \vert \psi_5^{\text{R}(1)}({\bm p})\rangle    
          - \langle\psi_5^{\text{L}(1)}({\bm p})\vert \Lambda({\bm p}) \vert \psi_5^{\text{R}(1)}({\bm p})\rangle  
               +  \langle\psi_5^{\text{L}(1)}({\bm p})\vert L_{\bm k}^{(1)}({\bm p}) \vert \psi_5^{\text{R}(0)}({\bm p})\rangle\ .
\label{eq:3.62a}
\ee
The first two terms on the right-hand side cancel by Eq.~(\ref{eq:3.59a}). To evaluate the third term we note that 
$L_{\bm k}^{(1)}({\bm p}) \vert \psi_5^{\text{R}(0)}({\bm p})\rangle = ik \vert (\hat{\bm k}\cdot{\bm p}) a_5({\bm p})\rangle$
has a component in the zero eigenvector space that is eliminated by the projection operator in Eq.~(\ref{eq:3.59a}).
This yields
\be
\omega_5^{(2)}({\bm k}) \big\langle\psi_5^{\text{L}(0)}({\bm p})\big\vert \psi_5^{\text{R}(0)}({\bm p})\big\rangle =
   -k^2 \big\langle(\hat{\bm k}\cdot{\bm p}) \psi_5^{\text{L}(0)}({\bm p})\big\vert\Lambda^{-1}({\bm p})\big\vert (\hat{\bm k}\cdot{\bm p}) \psi_5^{\text{L}(0)}({\bm p})\big\rangle\ .
\label{eq:3.62b}
\ee
Using the normalization (\ref{eq:3.54}) we finally obtain
\be
\omega_5^{(2)}({\bm k}) = D_T k^2\ ,
\label{eq:3.62c}
\ee
\ese
\end{widetext}
with the heat diffusion coefficient 
\be
D_T = \kappa/c_V
\label{eq:3.63}
\ee
The heat conductivity $\kappa$ is given by the same expression as in the SR case, Eq.~(\ref{eq:3.39}), but
the susceptibility in the Einstein relation (\ref{eq:3.63}) is now $c_V$ instead of $c_p$. 

\smallskip     
\paragraph{Plasmon damping and dispersion}
\label{par:III.B.2c}

An analogous analysis for the $\alpha=1,2$ channels confirms Eq.~(\ref{eq:3.52b}) for the eigenvalues at $O(k^0)$.
The eigenvalues at this order combine with the collision operator, and it is convenient to define
\be
\tilde\Lambda_{1,2} = \Lambda + \omega_{1,2}^{(0)}\ .
\label{eq:3.64}
\ee
At $O(k)$ we find contributions $\delta\psi_{1,2}^{\text{L,R}(1)}$ in the orthogonal space ${\cal L}_{\perp}$ that
need to be added to Eqs.~(\ref{eq:3.55c}) and (\ref{eq:3.56b}). These are
\bse
\label{eqs:3.65}
\bea
\big\langle\delta\psi_{1,2}^{\text{L}(1)}({\bm p})\big\vert &=& \big\langle\psi_{1,2}^{\text{L}(0)}({\bm p)}\big\vert\,L_{\bm k}^{(1)}({\bm p})\,
    \tilde\Lambda_{1,2}^{-1} \,{\cal P}_{\perp}\ ,
     \nonumber\\
\label{eq:3.65a}\\
\big\vert\delta\psi_{1,2}^{\text{R}(1)}({\bm p})\big\rangle &=&  {\cal P}_{\perp}
     \tilde\Lambda^{-1} L_{\bm k}^{(1)}({\bm p})\,\big\vert \psi_{1,2}^{\text{R}(0)}({\bm p})\big\rangle\ . 
      \nonumber\\
\label{eq:3.65b}
\eea
\ese
These expressions need to be interpreted as the solutions of the underlying integral equations, with uniqueness
enforced by the projection operator ${\cal P}_{\perp}$, see the comments in the context of Eq.~(\ref{eq:3.28a}).
There is no contribution to the eigenvalues at this order, but at $O(k^2)$ one finds a contribution in addition to Eq.~(\ref{eq:3.52c}):
\bse
\label{eqs:3.66}
\bea
\delta\omega_{1,2}^{(2)} &=& \frac{-k^2\,\omega_p^2}{2nm}\,\big\langle {\hat k}_i {\hat k}_j \sigma^{ij}({\bm p})\big\vert \tilde\Lambda_{1,2}^{-1}
     \big\vert {\hat k}_l {\hat k}_m \sigma^{lm}({\bm p})\big\rangle
\label{eq:3.66a}\\
&=&  \frac{-k^2\,\omega_p^2}{2nm}\,\frac{4}{3} \big\langle (\hat{\bm k}\cdot{\bm p})(\hat{\bm k}_{\perp}\cdot{\bm v}_p) \big\vert
      \tilde\Lambda_{1,2}^{-1} \big\vert (\hat{\bm k}\cdot{\bm p})(\hat{\bm k}_{\perp}\cdot{\bm v}_p)\big\rangle
      \nonumber\\
\label{eq:3.66b}
\eea
\ese
with $\sigma^{ij}({\bm p})$ from Eq.~(\ref{eq:3.43a}). In writing Eq.~(\ref{eq:3.66b}) we have used the fact that the
contribution $\sigma_2$ to the stress tensor, Eq.~(\ref{eq:3.43c}), vanishes for the model QP interaction we are using,
see the comment after Eq.~(\ref{eq:3.44b}). 

This contribution to the eigenfrequency has the structure of a transport coefficient; it can be interpreted as
a high-frequency shear viscosity, see Appendix E in Paper II. It has both a real part
and an imaginary part, so it contributes both to the plasmon dispersion and the plasmon damping. 
We emphasize that these results can {\em not} be obtained from the SR case by means of a simple substitution.
In a low-temperature expansion one can take advantage of the fact that the collision operator scales as some
positive power of the temperature and separate the real and imaginary parts of $\delta\omega_{1,2}^{(2)}$,
\begin{widetext}
\bse
\label{eqs:3.67}
\be
\omega_{1,2}({\bm k}) = \pm i \Omega_p({\bm k}) + \frac{1}{2}\,\Gamma_p\,k^2 + O(k^4)\ ,
\label{eq:3.67a}
\ee
Here
\be
\Omega_p({\bm k}) = \omega_p \sqrt{1 + c_1^2 k^2/\omega_p^2} + \frac{2}{3}\,\frac{1}{nm^*\omega_p}
     \Big\langle (\hat{\bm k}\cdot{\bm p})(\hat{\bm k}_{\perp}\cdot{\bm v}_p) \Big\vert
     1 + O(\Lambda^2) \Big\vert (\hat{\bm k}\cdot{\bm p})(\hat{\bm k}_{\perp}\cdot{\bm v}_p)\Big\rangle\, k^2 + O(k^4)   
\label{eq:3.67b}
\ee
is the wave-number dependent plasmon frequency, and
\be
\Gamma_p = \frac{-1}{nm^*\omega_p^2}\,\frac{4}{3}  \Big\langle (\hat{\bm k}\cdot{\bm p})(\hat{\bm k}_{\perp}\cdot{\bm v}_p) \Big\vert
     \Lambda + O(\Lambda^3) \Big\vert (\hat{\bm k}\cdot{\bm p})(\hat{\bm k}_{\perp}\cdot{\bm v}_p)\Big\rangle  
\label{eq:3.67c}
\ee
\ese
\end{widetext}
is the plasmon damping coefficient. 
Note that the leading contribution to the plasmon damping is linear in $\Lambda \sim 1/\tau(T)$, with 
$\tau(T)$ a relaxation time, and thus decreases with decreasing temperature, whereas
the first-sound damping is proportional to $\tau$ (see Eqs.~(\ref{eqs:3.44}) and (\ref{eq:3.34}), and thus increases with
decreasing temperature. A related observation is that the plasmon damping coefficient in the low-temperature limit
is no longer given by the solution of an integral equation, in contrast to the SR case, see Eqs.~(\ref{eqs:3.41}).
If we use the model BGK operator from Appendix~\ref{app:C} we find, to $O(k^2)$ and to lowest order in the temperature,
\bse
\label{eqs:3.68}
\bea
\Omega_p({\bm k}) &=& \omega_p + \frac{1}{2\omega_p} \,\left[c_1^2 + \frac{4}{15}\,(\vF^*)^2\right] k^2                
\label{eq:3.68a}\\
\Gamma_p &=& \frac{4}{15}\,\left(\frac{\vF^*}{\omega_p}\right)^2\,\frac{1}{\tau}\ .
\label{eq:3.68b}
\eea
\ese

\subsection{The fate of the hydrodynamic modes in the low-temperature limit}
\label{subsec:III.C}

As we have seen, the hydrodynamic modes in a degenerate Fermi liquid are essentially the same as those in a classical
fluid. With decreasing temperature the hydrodynamic regime shrinks; it is confined to wave numbers smaller than
\be
q^* \approx 1/\vF \tau(T)\ ,
\label{eq:3.69}
\ee
where $1/\tau(T)$ is a generic relaxation rate that vanishes as $T\to 0$.
At low temperatures for fixed wave number, or at larger wave numbers for fixed temperature, the system enters a
different regime where collisions between QPs are no longer dominating the physics and the soft modes are of a 
very different nature. It is sometimes said that the hydrodynamic modes ``cross over" to the collisionless ones. This
is misleading for various reasons. First, the hydrodynamic modes to not evolve into something else; rather, they
disappear since the regime that supports their existence shrinks to zero. Second, there is no one-to-one correspondence
between soft modes in the two regimes. The number of soft modes in the collisionless regime is not limited by the
number of conservation laws, and in some sense are infinitely many of them, as we will discuss next.

The above discussion pertains to the five hydrodynamic modes in the SR case, and to the modes that remain
soft in the LR case. The plasmon is qualitatively different: It is not a hydrodynamic mode, but rather a consequence of
gauge invariance.\cite{Anderson_1963} Therefore, it exists, and actually becomes better defined, even in the
absence of collisions. The plasmon excitation in the hydrodynamic regime is thus identical to the
one in collisionless regime, as we will see in Sec.~\ref{subsec:IV.B} below.

\section{Solutions of the kinetic equations II: Collisionless regime}
\label{sec:IV}
So far we have discussed the hydrodynamic regime, where the collision operator $\Lambda$ dominates the kinetic operator $L_{\bm k}$.
As we have seen, the soft modes in that case are controlled by the five conservation laws. The soft modes are the zero eigenfunctions
that correspond to the five zero eigenvalues of $\Lambda$, and their long-wavelength properties are given by the operator $L_{\bm k}^{(1)}$,
Eq.~(\ref{eq:3.4a}), which perturbs the zero eigenvalues. We now turn to the collisionless regime, which is defined by the linearized kinetic
operator $L_{\bm k}^{(1)}$ dominating over the collision operator. This realized for wave numbers larger than $q^*$ in Eq.~(\ref{eq:3.69}).
At $T=0$, where there are no collisions, the collisionless regime extends all the way to zero wave number. This regime has been discussed
extensively in the context of He-3,\cite{Abrikosov_Khalatnikov_1959, Baym_Pethick_1991} and we will focus on aspects that are either less
well known or important for other applications.

\subsection{Short-ranged case}
\label{subsec:IV.A}

Consider the kinetic equation (\ref{eq:2.16a}) at $T=0$, where $\Lambda({\bm p})=0$. Since the kinetic operator $L_{\bm k}$ is now
linear in ${\bm k}$, the entire distribution function $\phi$ is soft. Consequently, all of its moments with respect to the momentum ${\bm p}$
are soft, and in this sense we have an infinite number of soft modes; see Sec.~\ref{sec:V} for an elaboration.
For simplicity, let us first consider the case where the only nonzero Landau parameter
is $F_0$. Then Eq.~(\ref{eq:2.16a}) yields 
\be
\phi({\bm p},{\bm k},z) = \frac{F_0}{N_0}\,\frac{\hat{\bm p}\cdot\hat{\bm k}}{\zeta - \hat{\bm p}\cdot\hat{\bm k}}\,\delta n({\bm k},z)
   + \frac{i\,\phi({\bm p},{\bm k},t=0)}{z - \vF k\,( \hat{\bm p}\cdot\hat{\bm k})}\ ,
\label{eq:4.1}
\ee
where $\zeta = z/\vF k$. By summing over ${\bm p}$ we obtain a linear equation for the density fluctuation $\delta n$:
\bse
\label{eqs:4.2}
\be
\delta n({\bm k},z) \Bigl[1 - F_0 I_1(\zeta)\Bigr] = \frac{1}{V}\sum_{\bm p} w({\bm p})\,\frac{i\,\phi({\bm p},{\bm k},t=0)}{z - \vF k\, (\hat{\bm p}\cdot\hat{\bm k})}
\label{eq:4.2a}
\ee
where
\be
I_1(\zeta) = \frac{1}{2} \int_{-1}^{1} d\eta\,\frac{\eta}{\zeta - \eta} = -1 - \frac{\zeta}{2}\,\log\left(\frac{\zeta-1}{\zeta+1}\right)\ .
\label{eq:4.2b}
\ee
\ese
Substituting this expression back into Eq.~(\ref{eq:4.1}) yields an explicit expression for $\phi$. We see that all of the soft
modes are characterized by a propagator
\be
P({\bm k},z) = P(\zeta) = \frac{1}{1 - F_0 I_1(\zeta)} - 1\ .
\label{eq:4.3}
\ee
Here the constant subtraction term serves to make $P$ a proper causal function that vanishes for $\zeta\to\infty$. 
The spectrum of $P$, $P''({\bm k},\omega) = \Im P({\bm k},\omega + i0)$,  consists of a continuous part and, in addition, 
delta-function contributions that correspond to zeros of the denominator. Two such zeros exist for any $F_0>0$; 
the are the well known zero-sound modes.\cite{Landau_Lifshitz_IX_1991} The resonance frequencies (which correspond to
$-i$ times the perturbed zero eigenvalues $\omega$ in Sec.~\ref{sec:III}) are
\bse
\label{eqs:4.4}
\be
z = \pm c_0 k\ ,
\label{eq:4.4a}
\ee
where the speed of zero sound is given by
\be
c_0 = \sigma_0 \vF
\label{eq:4.4b}
\ee
with $\sigma_0$ is the solution of
\be
I_1(\sigma_0) = 1/F_0\ ,
\label{eq:4.4c}
\ee
\ese
which exists and is unique for all $F_0>0$.

If we add the Landau parameter $F_1$, we obtain an expression for $\phi$ in terms of the density fluctuation $\delta n$ and the velocity
fluctuation $\delta{\bm u}$ that is a generalization of Eq.~(\ref{eq:4.1}): 
\begin{widetext}
\be
\phi({\bm p},{\bm k},z) = \frac{F_0}{N_0}\,\frac{\hat{\bm p}\cdot\hat{\bm k}}{\zeta - \hat{\bm p}\cdot\hat{\bm k}}\,\delta n({\bm k},z)
     + \frac{F_1}{N_0}\,\frac{n}{\vF}\,\frac{\hat{\bm p}\cdot\hat{\bm k}}{\zeta - \hat{\bm p}\cdot\hat{\bm k}}\,\hat{\bm p}\cdot\delta{\bm u}({\bm k},z)
        + \frac{i\,\phi({\bm p},{\bm k},t=0)}{z - \vF k\, \hat{\bm p}\cdot\hat{\bm k}}\ .
\label{eq:4.5}
\ee
\end{widetext}

Using this expression to calculate $\delta n$ and $\delta{\bm u}$ via Eqs.~(\ref{eq:2.17a}, \ref{eq:2.17b}) we obtain a 2$\times$2
system for $\delta n$ and $\delta u_L$ whose determinant generalizes the propagator $P$ to
\bse
\label{eqs:4.6}
\be
P_L(\zeta) = \frac{1}{1+F_1/3 - \left[F_0(1+F_1/3) + F_1\zeta^2\right] I_1(\zeta)} - 1\ .
\label{eq:4.6a}
\ee
The spectrum is qualitatively the same as for $F_1=0$.
In addition, the equation for the transverse velocity yields a transverse propagator
\be
P_T(\zeta) = \frac{1}{1 - F_1/6 + \frac{1}{2}\,(\zeta^2-1) I_1(\zeta)}
\label{eq:4.6b}
\ee
\ese 
The spectrum of $P_T$ also has a continuous part, and for $F_1>6$ it contains, in addition, transverse zero-sound modes.

\begin{figure}[b]
\vskip 0pt
\includegraphics[width=8cm]{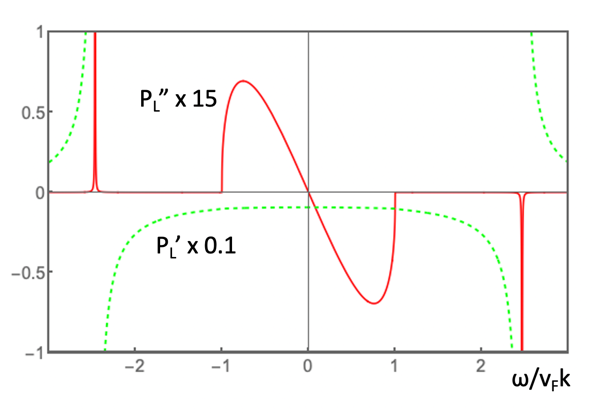}
\caption{Spectrum $P_L"$ (solid red) and reactive part $P_L\!'$ (dashed green) of the longitudinal propagator, Eq.~(\ref{eq:4.6a}), for $F_0 = F_1 = 5$.
              The spectrum consists of the continuous unparticle contribution in the center and the two zero-sound
              delta-function contributions. The transverse propagator, Eq.~(\ref{eq:4.6b}), is qualitatively the same except that the zero-sound contributions
              to the spectrum exist only if $F_1>6$. For either propagator the entire spectrum is scale invariant.}
\label{fig:2}
\end{figure}

These results demonstrate that the soft modes in the collisionless regime are qualitatively different from those in the
hydrodynamic regime: There always is a continuous part of the spectrum, which represents a continuum of soft modes
that obey a linear scaling of the frequency with the wave number. This continuous part is not particle-like and has no 
analog in the hydrodynamic regime; in a particle-physics context such continuous scale invariant excitations have
been dubbed `unparticles'.\cite{Georgi_2007} In addition, there are particle-like zero-sound modes that are unrelated to 
conservation laws; how many of these there are depends on the QP interaction.
Figure~\ref{fig:2} demonstrates these features. We emphasize that the continuum and the zero-sound poles 
are part of the same spectrum, but the continuum is the more fundamental part in the sense that it does not depend on
the values of the Landau parameters and exists even in a noninteracting Fermi system. 
%
%
%
%\begin{figure}[h]
%\vskip 20pt
%\includegraphics[width=8cm]{P_T_plot.png}
%\caption{Spectrum and reactive part of the transverse propagator for $F_0 = F_1 = 5$.}
%\label{fig:3}
%\end{figure}
%
%
%\begin{figure}[h]
%\includegraphics[width=8cm]{P_T_2_plot.png}
%\caption{Spectrum and reactive part of the transverse propagator for $F_0$, $F_1 = 15$.}
%\label{fig:3}
%\end{figure}
%

So far we considered only the Landau parameters $F_0$ and $F_1$. If one keeps higher terms in the
expansion of $F({\bm p},{\bm p}')$ in Legendre polynomials, Eq.~(\ref{eq:2.8}), additional zero-sound
modes can appear. To demonstrate this, we consider a model that keeps the Landau parameter $F_2$
and repeat the above analysis. The longitudinal propagator then becomes
\begin{widetext}
\bse
\label{eqs:4.7}
\be
P_L(\zeta) = \frac{1}{(1+\frac{1}{3}F_1)\left[1 - \frac{1}{20}(1+5 F_0) F_2\right] + \frac{3}{4} F_2 \zeta^2 - d(\zeta) I_1(\zeta)} - 1\ ,
\label{eq:4.7a}
\ee
where
\be
d(\zeta) =  (1+\frac{1}{3}F_1)\left[F_0 + \frac{1}{4}(1+\frac{9}{5}F_0)F_2\right] 
     + \left[ F_1 - \frac{3}{2}\left(1 + \frac{1}{2} F_0(1 + \frac{1}{3} F_1) + \frac{1}{30} F_1\right)F_2 \right] \zeta^2 + \frac{9}{4} F_2 \zeta^4
\label{eq:4.7b}
\ee
\ese
\end{widetext}
An inspection shows that the denominator of $P_L$ has one zero that is continuously related to the zero-sound pole 
in Eq.~(\ref{eq:4.6a}). It has a second zero, closer to the continuous part of the spectrum, provided the following
conditions are fulfilled:
\bse
\label{eqs:4.8}
\be
F_0 >  \frac{10}{3}\,\frac{1 + F_1/30}{1 + F_1/3}
\label{eq:4.8a}
\ee
and
\be
F_2 > \frac{30 F_0(1+\frac{1}{3} F_1)}{9 F_0(1+\frac{1}{3}F_1) - F_1 - 30}\ .
\label{eq:4.8b}
\ee
\ese
While the existence of the additional collective mode requires a rather large positive value of $F_2$, 
it illustrates again that the zero-sound modes are governed by the QP interaction rather than by
conservation laws.

At a low nonzero temperature the collision operator gives all of the soft modes in the collisionless regime a mass. 
For the zero-sound modes this takes the form of a damping term that broadens the $\delta$-function in the spectrum
of the propagator. For an explicit example we use the BGK model collision operator from Appendix~\ref{app:C}, with
$\tau(T)$ the temperature-dependent relaxation time. The
infinitely many soft modes now couple, and the kinetic equation is no longer exactly soluble even if one keeps only
the Landau parameter $F_0$. In an approximation that keeps only the density and the longitudinal momentum,
and ignores the coupling to the modes at higher angular momenta, we find for the resonance frequencies that
generalize Eq.~(\ref{eq:4.4a}) to linear order in $1/\tau$
\bse
\label{eq:4.9}
\be
z = \pm c_0 k - i\gamma\ ,
\label{eq:4.9a}
\ee
with $c_0$ from Eq.~(\ref{eq:4.4b}) and
\be
\gamma = \frac{1}{\tau(T)} \left[ 1 - \frac{1 + F_0 + 3\sigma_0^2}{F_0^2 \sigma_0 \vert I_1'(\sigma_0)\vert}\right]\ ,
\label{eq:4.9b}
\ee
\ese
which is positive for all $F_0>0$.
Here $I_1'$ is the derivative of the function $I_1$ from Eq.~(\ref{eq:4.2b}). Note that the damping coefficient
$\gamma$ is independent of the wave number, i.e., the mode is massive for all $T>0$. $\gamma$ vanishes at $T=0$,
and increases with increasing temperature. This is qualitatively different from the damping of the first-sound
mode in the hydrodynamic regime, Eq.~(\ref{eq:3.41a}), which is proportional to $k^2$ and increases with
decreasing temperature.

The damping of other modes can be analyzed analogously. In particular, the continuum unparticle excitation
acquires a mass that is proportional to $1/\tau$.

\subsection{Long-ranged case}
\label{subsec:IV.B}

\begin{figure}[b]
\includegraphics[width=8cm]{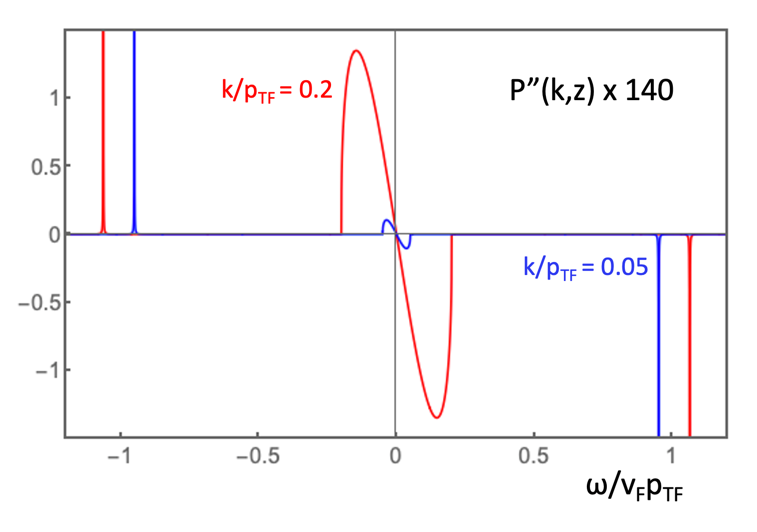}
\caption{Spectrum of the propagator in Eq.~(\ref{eq:4.10}) for $F_0 = 5$ and $k=0.2\,p_{\text{TF}}$ (red) and $k=0.05\,p_{\text{TF}}$ (blue),
              respectively, showing the continuous unparticle conribution and the plasmon poles. The continuum is still scale 
              invariant, with  the frequency scaling linearly with the wave number, whereas the plasmon poles are not and depend 
              only weakly on the wave number.}
\label{fig:3}
\end{figure}

In the collisionless regime we can deduce the spectrum in the LR case by substituting Eq.~(\ref{eq:2.11})
into the results for the SR case. We illustrate the result by performing this substitution in the propagator
$P$ from Eq.~(\ref{eq:4.3}), which now reads
\be
P({\bm k},z) = \frac{1}{1 - [F_0 + p_{\text{TF}}^2(1+F_0)/k^2] I_1(\zeta)} - 1\ .
\label{eq:4.10}
\ee
The continuous part of the spectrum is qualitatively the same as in the SR case, but instead of the
zero-sound pole we now have a plasmon pole at frequency
\be
z = \pm\omega_p + O(k^2)
\label{eq:4.11}
\ee
The continuous excitation still displays scale invariance, but the plasmon does not. This is illustrated
in Fig.~\ref{fig:3}, which shows the spectrum for two difference wave numbers. 

The dispersion of the plasmon and its damping can be obtained by keeping terms to $O(k^2)$.
A calculation analogous to the one that leads to Eq.~(\ref{eq:4.9b}) yields
\bse
\label{eqs:4.12}
\bea
z &=&  \pm\omega_p \left[1 + \frac{1}{2}\left(\frac{3}{5} + \frac{1}{3}\,F_0\right)\left(\frac{\vF^*}{\omega_p}\right)^2 k^2\right] - \frac{i}{2}\Gamma_p\,k^2 
\nonumber\\
&& \hskip 120pt  + O(k^4)\ ,
\label{eq:4.12a}
\eea
with
\be
\Gamma_p = \frac{4}{15}\,(\vF^*/\omega_p)^2\,\frac{1}{\tau(T)}
\label{eq:4.12b}
\ee
\ese
the plasmon damping coefficient. This is the same result as in the hydrodynamic regime, see
Eqs.~(\ref{eqs:3.68}).

We emphasize that the only zero-sound mode that becomes massive in the LR case
is the one that is present even with $F_0$ the only nonzero
Landau parameter. For instance, the additional soft mode discussed
in connection with Eqs.~(\ref{eqs:4.7}) remains soft, so does the transverse zero-sound mode described 
by Eq.~(\ref{eq:4.6b}), and so does the continuous unparticle excitation.

\subsection{The fate of the collisionless modes with increasing temperature}
\label{subsec:IV.C}

The collisionless regime is complementary to the hydrodynamic one, it is confined to wave numbers
larger than $q^*$ defined in Eq.~(\ref{eq:3.69}). The collisionless soft modes are truly soft only at
$T=0$; for any nonzero temperature they acquire a damping term that does not vanish in the limit
of zero wave number, but still is small for low temperatures. With increasing temperature the relaxation
time decreases, and the collisionless regime gets pushed to larger wave numbers and frequencies,
while the hydrodynamic regime grows. At the same time, the damping of the modes increases and they 
eventually become overdamped, while the hydrodynamic modes emerge at low frequencies and
wave numbers and their damping decreases.

As already mentioned in Sec.~\ref{subsec:III.C}, these observations do not apply to the plasmon,
which is governed by gauge invariance and therefore is the same in both the hydrodynamic
and the collisionless regimes, see Eqs.~(\ref{eqs:3.68}) and (\ref{eqs:4.12}).

\section{Discussion and Conclusion}
\label{sec:V}

We conclude with a summary and discussion of some of the salient points of the paper. We also add some remarks
regarding points that were mentioned only briefly, or not at all, in the main text.

\begin{enumerate}[label=\arabic*), labelwidth=!, labelindent=9pt, leftmargin=0pt, itemindent=0pt]
\item
We have addressed two very general questions regarding Fermi liquids: First, we have shown that LFL theory, which is often thought of as
being valid only at low temperatures, is fully consistent with Navier-Stokes hydrodynamics irrespective of the
temperature.
%We have shown this explicitly for
We have done so by using kinetic theory to explicitly solve the kinetic equation for the
hydrodynamic modes, by means of the method of perturbed zero eigenvalues of the collision operator. 
Alternatively, one can derive the Navier-Stokes equations for the LFL. This program will be carried out in Paper II.
For our explicit solution we have used
the model interaction given by Eq.~(\ref{eq:2.8}), which implies (\ref{eq:2.10a}).
However, we have used Eq.~(\ref{eq:2.10a}) only twice: Once to eliminate the additional term on the right-hand-side
of Eq.~(\ref{eq:3.42a}), and once to ascertain that the model kinetic operator $L_{\bm k}^{(1)}({\bm p})$ is consistent with 
particle number conservation. This, and the general structure of the theory, strongly suggests that an analogous analysis is 
possible for a completely general QP interaction. Carrying out this program will lead to a qualitatively new effect, namely, a
component of the QP velocity in the orthogonal space ${\cal L}_{\perp}$. That is, physical particles and quasiparticles
will have the same density, but different currents. The resulting hydrodynamic theory will have a structure that is
different from that of a simple classical fluid and share some (but not all) aspects with a classical binary mixture. Among
the physical consequences will be a nonzero bulk viscosity, and an additional contribution to the sound attenuation.

It should be mentioned that the fact that LFL theory is internally consistent, and consistent with general hydrodynamics,
at all $T$ does not imply that it LFL theory is. For instance, nonlocalities in the collision operator (i.e., different
single-particle distributions occurring at different points in real space) will lead to contributions at $O(T^2)$ that are
not included in LFL theory.

\hskip 10pt Second, we have discussed the absence of a relation between the soft modes in the hydrodynamic and collision regimes,
respectively. With decreasing temperature the damping of the hydrodynamic modes increases and the hydrodynamic regime
shrinks until it disappears at $T=0$. At the same time, a completely unrelated family of soft modes emerges in the collisionless
regime. Their number is governed by the QP interaction rather than by conservation laws, their damping decreases with
decreasing temperature, and they are truly soft only at $T=0$. Tables~\ref{table:1} and \ref{table:2} summarizes the soft modes, as well as
the plasmon modes, in both regimes.

\hskip 10pt We have also discussed the special role played by the plasmon in a charged Fermi liquid, which is not a hydrodynamic mode
and extends through both the collisionless and the hydrodynamic regimes for reasons related to gauge invariance.\cite{Anderson_1963}
We have discussed 3-d systems where the plasmon is massive and its damping is independent of the wave number
in the homogeneous limit, see Eqs.~(\ref{eqs:3.68}) and (\ref{eqs:4.12}). This changes in 2-d systems, where both the
plasmon frequency and the damping go to zero as $k\to 0$; the former as $\sqrt{k}$ and the latter as $k^2$.\cite{Lucas_Das_Sarma_2018}

\hskip 10pt We also note that `Fermi liquid', in our context, can be interpreted rather broadly: We have not specified the temperature
dependence of the relaxation rate $1/\tau$, and we have not made use of the concept of `well-defined quasiparticles'.
For instance, our analysis of the hydrodynamic regime applies to what is known as a marginal Fermi liquid.\cite{Varma_et_al_1989}

\begin{table}[b]
\caption{Modes in the hydrodynamic regime}
%\smallskip
\begin{ruledtabular}
\begin{tabular}{lll}
                              & Short Ranged                      &  Long Ranged \\
\hline\\                              
Shear modes (2)   & diffusive                                &  diffusive        \\
                             & $z \sim -iD_{\perp}k^2$       &  $z \sim -iD_{\perp}k^2$ \\
\\
Heat mode (1)     & diffusive                                &  diffusive        \\
                             & $z \sim -iD_Tk^2$                &  $z \sim -iD_Tk^2$     \\
\\
First sound (2)   & propagating                             &  none              \\    
                          & $z\sim c_1 k - i\vF k^2\tau$ &      \\    
\\
Plasmon (2)      &  none                                       &  propagating   \\   
                         &                                                 &  $z\sim \omega_p - i\left(\frac{\vF k}{\omega_p}\right)^2 \frac{1}{\tau}$  \\
\end{tabular}
\end{ruledtabular}
\vskip 0pt
\label{table:1}
\end{table}
\begin{table}[h]
\caption{Modes in the collisionless regime}
%\smallskip
\begin{ruledtabular}
\begin{tabular}{lll}
                              & Short Ranged   &  Long Ranged \\
\hline\\      
Unparticle (1)               &  continuous function          & continuous function  \\  
                                    &  of  $z/\vF k$                      &  of $z/\vF k$               \\
\\                                    
$\ell=0$ zero               & propagating                        &  none        \\
sound (2)                    & $z \sim c_0 k - i/\tau$         &                         \\
\\
Plasmon (2)               & none                                    &  propagating       \\
                                   &                                            &  $z\sim \omega_p - i\left(\frac{\vF k}{\omega_p}\right)^2 \frac{1}{\tau}$     \\
\\
$\ell>0$ zero             & propagating                         &  propagating                            \\    
sound (many)           & $z \sim c_0 k - i/\tau\ ^{*)}$          &   $z \sim c_0 k - i/\tau\ ^{*)}$            \\  [5pt]  
\hline\\[-5pt]
\multicolumn{3}{l} {$^{*)}$ The zero sound velocities and the damping coefficients are} \\
\multicolumn{3}{l}{different for different modes. However, they all are propor-} \\
\multicolumn{3}{l}{tional to those of the $\ell=0$ mode given in Eqs.~(\ref{eq:4.4b}) and } \\
\multicolumn{3}{l}{(\ref{eq:4.9b}), respectively.}
\end{tabular}
\end{ruledtabular}
%\vskip -5mm
\label{table:2}
\end{table}

\item
The number of soft modes in the collisionless regime (or, strictly speaking, at $T=0$), is to some extent a matter of
interpretation. In the ${\bm p}\,$-${\bm k}$ momentum space, sometimes referred to as $\mu$-space in kinetic theory,\cite{Dorfman_vanBeijeren_Kirkpatrick_2021}
there is only one soft mode, viz., the fluctuation $\phi$ of the single-particle distribution function, whose denominator
is given by $z - {\bm p}\cdot{\bm k}$, see Eq.~(\ref{eq:4.1}). However, as a result of this denominator all of the
moments of $\phi$ with respect to ${\bm p}$ are soft, and in this sense there is an infinite number of soft modes. This 
is true in a clean Fermi system; in the presence of quenched disorder only the zeroth moment with respect to ${\bm p}$
is soft, see Refs.~\onlinecite{Wegner_1979, Schaefer_Wegner_1980, Finkelstein_1983, Belitz_Kirkpatrick_1997,
Belitz_Evers_Kirkpatrick_1998}, for a review see Ref.~\onlinecite{Belitz_Kirkpatrick_1994}.

\hskip 10pt It should be emphasized that the spectrum of $\phi$ has a continuous part and, in general, $\delta$-function contributions
describing zero-sound modes that are both part of the same spectrum (this supports the single-soft mode interpretation). 
Our results are consistent with a quantum-field-theoretic analysis in Ref.~\onlinecite{Belitz_Kirkpatrick_2012a}, which kept only
the equivalent of the Landau parameter $F_0$. It is interesting that kinetic theory, which uses quantum mechanics only
in the form of the equilibrium Fermi distribution, and the field theory are equivalent. Following Ref.~\onlinecite{Georgi_2007}, 
we have referred to the continuous part of the spectrum as the unparticle excitation, whereas the zero-sound poles represent 
particle-like excitations. We note that the unparticle part of the spectrum, while an exotic idea in a high-energy context, has been 
known since the earliest days of the quantum theory of condensed matter, where it is usually referred to as the particle-hole
continuum. For instance, it gives the Lindhard function\cite{Lindhard_1954} its characteristic scale-invariant structure.
Neither the continuum nor the zero-sound poles are related to conservation laws, and all of them acquire a mass
at any nonzero temperature. Some remarks to the contrary related to zero sound in Ref.~\onlinecite{Belitz_Kirkpatrick_2012a}
were incorrect.

\hskip 10pt We have discussed only spinless Fermi liquids for simplicity's sake. It is well known how to incorporate spin in
LFL theory,\cite{Landau_Lifshitz_IX_1991, Baym_Pethick_1991} and the generalization in the current context is
straightforward. The spin channel again supports the unparticle continuum, and in addition spin-zero-sound modes
whose existence and number depends on the values of the Landau parameters.

\hskip 10pt The importance of the unparticle continuum is often downplayed in favor of the particle-like collective zero-sound
excitations. This ignores the fact that it has dramatic physical consequences. For instance, in the spin channel (which
we have not explicitly discussed) it is responsible for a nonanalytic wave-number dependence of the spin 
susceptibility,\cite{Belitz_Kirkpatrick_Vojta_1997} and for the ferromagnetic quantum phase transition to be generically 
a first-order transition.\cite{Belitz_Kirkpatrick_Vojta_1999, Brando_et_al_2016a}

\item
The origin and interpretation of the soft modes in the collisionless regime has been the subject of several studies.
The scale invariant unparticle continuum mode has been interpreted as the Goldstone mode of a spontaneously broken 
rotational symmetry in Matsubara frequency space; roughly speaking, a broken symmetry between retarded and
advanced degrees of freedom.\cite{Belitz_Kirkpatrick_2012a} This is in analogy to Wegner's interpretation of the
diffusive soft mode in disordered Fermi systems known as the `diffuson'.\cite{Wegner_1979} Reference~\onlinecite{Alberte_Nicolis_2020}
has interpreted it as a Goldstone mode related to a spontaneously broken Lorentz boost invariance. The relation
between these two interpretations is not clear.

\item
We have discussed clean fermion systems, but impurity scattering can easily be taken into account; see
Appendix~\ref{app:C} for the relevant collision integral. It qualitatively changes the hydrodynamic modes:
fermion momentum is no longer conserved, and the density response is diffusive. In bulk metals the clean 
hydrodynamic behavior we have discussed is very difficult to realize, since impurity scattering tends to 
dominate even in the cleanest samples. In two-dimensional systems the ultraclean hydrodynamic 
regime, where momentum is conserved, is easier to realize.\cite{Briskot_et_al_2015, Lucas_Das_Sarma_2018}
Also, it recently has become possible to realize clean Fermi liquids in cold-atom systems.
\cite{Cao_et_al_2011, Bloch_Dalibard_Nascimbene_2012, Joseph_Elliott_Thomas_2015, Bluhm_Hou_Schaefer_2017, 
Patel_et_al_2020, Wang_et_al_2021}

\item
We have discussed LFL theory at a level analogous to the linearized Navier-Stokes equations of classical
hydrodynamics. An interesting problem is the generalization of this treatment by adding a fluctuating Langevin
force. This will be analogous to the fluctuating hydrodynamics description of classical fluids\cite{Landau_Lifshitz_VI_1959}
and allow for the calculation of time-correlation functions in both equilibrium and non-equilibrium situations. 
This problem is considered in Paper II.\cite{Kirkpatrick_Belitz_2022}

\end{enumerate}

\appendix
\section{Thermodynamic relations}
\label{app:A}

Here we list and explain various thermodynamic relations that were used in Secs.~\ref{sec:II} and \ref{sec:III}.
We start with the normalization factors in Eqs.~(\ref{eqs:2.21}). 

\subsection{Normalizations of the zero eigenvectors}
\label{app:A.1}

To determine $A_1$, consider the variation $\delta f_{\text{eq}}({\bm p})$ of the equilibrium distribution function due to a homogeneous 
variation $\delta\mu$ of the chemical potential at fixed $T$ and $V$:
\be
\delta f_{\text{eq}}({\bm p})\Bigr\vert_{T,V} = -w({\bm p})\left(\delta\epsilon_p - \delta\mu\right)\ .
\label{eq:A.1}
\ee
$\delta\epsilon_p$ is related to $\delta f_{\text{eq}}$ by Eq.~(\ref{eq:2.6}), and by symmetry only the Landau parameter $F_0$ contributes,
\be
\delta\epsilon_p = \frac{1}{N_0}\,F_0\,\delta n\ .
\label{eq:A.2}
\ee
For the variation of the number density, Eq.~(\ref{eq:2.1b}), we thus have
\be
\delta n = \frac{N_0}{1+F_0}\,\delta\mu
\label{eq:A.3}
\ee
with $N_0 = \langle 1\vert 1\rangle$ from Eq.~(\ref{eq:2.4c}). Hence, \cite{Baym_Pethick_1991, Pines_Nozieres_1989}
\be
\left(\frac{\partial n}{\partial\mu}\right)_{T,V} = \frac{\langle 1\vert 1\rangle}{1+F_0}\ ,
\label{eq:A.4}
\ee
which is the second equality in Eq.~(\ref{eq:2.21b}). In the limit of low temperature,
\be
\langle 1\vert 1\rangle = \NF^* + O(T^2)\ .
\label{eq:A.5}
\ee

The same line of reasoning for a variation of $f_{\text{eq}}$ under a variation $\delta T$ at fixed $\mu$ and $V$ yields
\be
\left(\frac{\partial n}{\partial T}\right)_{\mu,V} = \frac{1}{1+F_0}\,\frac{1}{T}\,\langle\xi_p\vert 1\rangle\ .
\label{eq:A.6}
\ee
Combining this with Eq.~(\ref{eq:A.4}) we obtain
\be
\langle \xi_p\rangle_w = \langle\epsilon_p\rangle_w - \mu = - T(\partial\mu/\partial T)_{N,V}\ .
\label{eq:A.7}
\ee
From Eq.~(\ref{eq:A.7}) we can obtain the normalization $A_3$ as follows. Consider a variation of the energy density, Eq.~(\ref{eq:2.1a}).
For fixed number density $n$ we have $\delta\epsilon_p = 0$, and hence
\bse
\label{eqs:A.8}
\be
\delta f_{\text{eq}}({\bm p})\Bigr\vert_{N,V} = w({\bm p})\left(\delta\mu + \frac{1}{T}\,\xi_p\,\delta T\right)\ ,
\label{eq:A.8a}
\ee
and therefore, from Eq.~(\ref{eq:2.1a}),
\be
\delta e = \langle\epsilon_p\vert 1\rangle \delta\mu + \frac{1}{T}\,\langle\epsilon_p\vert\xi_p\rangle \delta T\ .
\label{eq:A.8b}
\ee
\ese
For the specific heat at constant volume this yields
\be
T c_V = T \left(\frac{\partial e}{\partial T}\right)_{V,N} = T \langle\epsilon_p\vert 1\rangle \left(\frac{\partial\mu}{\partial T}\right)_{V,N} + \langle\epsilon_p\vert\xi_p\rangle\ .
\label{eq:A.9}
\ee
By using Eq.~(\ref{eq:A.7}) we obtain
\be
T c_V = \langle\epsilon_p\vert\epsilon_p\rangle - \langle\epsilon_p\rangle_w^2 \langle 1\vert 1\rangle = \langle a_5({\bm p})\vert a_5({\bm p})\rangle
\label{eq:A.10}
\ee
with $a_5$ from Eq.~(\ref{eq:2.20e}). This is the second equality in Eq.~(\ref{eq:2.21d}). Alternatively, we obtain the same result by starting
with the entropy density of a Fermi liquid in the form\cite{Landau_Lifshitz_IX_1991, Baym_Pethick_1991}
\be
s = \frac{-1}{V}\sum_{\bm p}\Bigl[ f_{\text{eq}} \ln f_{\text{eq}} + (1 - f_{\text{eq}})\ln(1 - f_{\text{eq}})\Bigr] \ ,
\label{eq:A.11}
\ee
using Eq.~(\ref{eq:A.8a}), and calculating the specific heat as $c_v = T(\partial s/\partial T)_{V,N}$.

For $A_2$, we need the same expression as for $\langle 1\vert 1\rangle$ with an additional ${\bm p}^2$ in the integrand. 
At $T=0$, the ${\bm p}^2$ gets replaced by $\pF^2$, and by using Eq.~(\ref{eq:A.5}) we have
$\langle{\bm p}\vert{\bm p}\rangle = \pF^2 \NF^* + O(T^2)$. Due to the f-sum rule this remains valid in general, i.e., there 
are no explicit temperature corrections,\cite{Pines_Nozieres_1989} and we have
\be
\langle{\bm p}\vert{\bm p}\rangle = \pF^2 \NF^* = 3 m^* n \ .
\label{eq:A.12}
\ee

We finally list some of the above quantities explicitly in the low-temperature limit. 
The normalizations are, in addition to Eq.~(\ref{eq:A.12}), which is valid at all $T$,
\bse
\label{eqs:A.12.1}
\bea
\langle 1 \vert 1 \rangle &=& \NF^* + O(T^2)\ ,
\label{eq:A.12.1a}\\
\langle a_5({\bm p})\vert a_5({\bm p})\rangle &=& c_v T = sT + O(T^4) 
\nonumber\\
   &=& \frac{\pi^2}{3} \NF^*\, T^2 + O(T^4)\ .
\label{eq:A.12.1b}
\eea
\ese
In Eq.~(\ref{eq:A.12.1b}) we have used the LFL theory expression for the entropy at low temperature.\cite{Baym_Pethick_1991}
Also of interest is $\langle\epsilon_p\rangle_w$. The chemical potential is\cite{Baym_Pethick_1991}
\be
\mu(T) = \epsilonF - \frac{\pi^2}{12}\,\frac{T^2}{\mu(T=0)} + O(T^4)
\label{eq:A.13}
\ee
Equation~(\ref{eq:A.7}) thus yields
\bea
\langle\epsilon_p\rangle_w &=& \mu + \frac{\pi^2}{6}\,\frac{T^2}{\mu} + O(T^4)
\nonumber\\
&=& \epsilonF + \frac{\pi^2}{12}\,\frac{T^2}{\mu} + O(T^4)\ .
\label{eq:A.14}
\eea

\subsection{Temperature fluctuations}
\label{app:A.2}

In order to derive Eq.~(\ref{eq:2.19f}) we start with Eqs.~(\ref{eq:A.1}) and (\ref{eq:A.2}) to write energy density fluctuations at constant $T$
and $V$ as
\bea
\delta e\vert_{T,V} &=& \frac{-1}{V} \sum_{\bm p} \epsilon_p w({\bm p})(\delta\epsilon_p - \delta\mu) 
\nonumber\\
&=& \left(\delta\mu - \frac{F_0}{\langle 1\vert 1\rangle} \delta n\right) \langle\epsilon_p\vert 1\rangle
\label{eq:A.16}
\eea
With the help of Eq.~(\ref{eq:A.4}) this yields
\be
(\partial e/\partial n)_{T,V} = - c_V(\partial T/\partial n)_{E,V} = \langle\epsilon_p\rangle_w\ .
\label{eq:A.17}
\ee
The first equality is generally valid, the second one is valid within LFL  theory. Now consider
\bea
c_V \delta T &=& c_V \left(\frac{\partial T}{\partial e}\right)_{N,V} \delta e + c_V \left(\frac{\partial T}{\partial n}\right)_{E,V} \delta n
\nonumber\\
&=& \delta e - \langle\epsilon_p\rangle_w \delta n = \langle a_5\vert\phi\rangle\ ,
\label{eq:A.18}
\eea
which is Eq.~(\ref{eq:2.19f}). Here we have used $c_V = (\partial e/\partial T)_{V,N}$ and Eq.~(\ref{eq:A.17}) to go from the
first line to the second one.

\subsection{The matrix $L_L^{(1)}$}
\label{app:A.3}

Equation~(\ref{eq:3.12b}) can be obtained from Eq.~(\ref{eq:3.12a}) by means of the following manipulations. 

In order to calculate $\langle {\bm v}_p\vert{\bm p}\rangle$, consider
\bse
\label{eqs:A.19}
\bea
\langle v_p^i \vert p_j\rangle &=& \frac{1}{V} \sum_{\bm p} w({\bm p})\,p_j\,\frac{\partial}{\partial p_i}\epsilon_p = \frac{-1}{V}\sum_{\bm p} p_j\,\frac{\partial f_{\text{eq}}}{\partial p_i} 
\nonumber\\
&=& \delta_{ij}\,\frac{1}{V} \sum_{\bm p} f_{\text{eq}}({\bm p}) = \delta_{ij}\,n\ ,
\label{eq:A.19a}
\eea
where the second line is obtained from the first one by partial integration. We thus have
\be
\langle {\bm v}_p \vert {\bm p}\rangle = 3n\ .
\label{eq:A.19b}
\ee
\ese
This yields the (1,2) matrix element in Eq.~(\ref{eq:3.12b}). 

$\partial n/\partial\mu$ is related to the compressibility $\chi$:
\bse
\label{eqs:A.20}
\be
(\partial n/\partial\mu)_{T,V} = n^2\chi_T\ ,
\label{eq:A.20a}
\ee
where
\be
\chi_T = \frac{-1}{V} \left(\frac{\partial V}{\partial p}\right)_{T,N} = \frac{1}{n} \left(\frac{\partial n}{\partial p}\right)_{T,V}\ .
\label{eq:A.20b}
\ee
\ese
is the isothermal compressibility, with $p$ the pressure. These identities follow from standard Jacobian manipulations\cite{Landau_Lifshitz_V_1980} 
combined with the fact that the particle number is an extensive quantity, and hence $(\partial N/\partial V)_{p,T} = N/V = n$. Combining 
Eqs.~(\ref{eq:A.20b}) and (\ref{eq:A.19b}) we obtain the (2,1) matrix element in Eq.~(\ref{eq:3.12b}).

To express $\langle {\bm v}_p\vert{\bm p}\,a_5({\bm p})\rangle$ in terms of thermodynamic quantities, we start with
\bea
\langle v_p^i \vert p_j\,\epsilon_p\rangle &=& \delta_{ij}\,\frac{1}{V}\sum_{\bm p} \epsilon_p\,f_{\text{eq}}({\bm p}) + \frac{1}{V}\sum_{\bm p} f_{\text{eq}}({\bm p})\,p_j\,v_p^i
\nonumber\\
&=& \delta_{ij}\,\frac{1}{V}\sum_{\bm p} \epsilon_p\,f_{\text{eq}}({\bm p}) + \delta_{ij}\,\frac{T}{V} \sum_{\bm p} \ln(1-f_{\text{eq}})\ .
\nonumber\\
\label{eq:A.21}
\eea
For the first line we have integrated by parts as in Eq.~(\ref{eq:A.19a}), and to obtain the second line we have used the identity from the
first line of Eq.~(\ref{eq:2.3b}). The entropy, Eq.~(\ref{eq:A.11}), can be rewritten as
\be
s = \frac{-1}{V}\sum_{\bm p} \ln(1-f_{\text{eq}}) + \frac{1}{TV}\sum_{\bm p} \epsilon_p\,f_{\text{eq}}({\bm p}) - n\mu/T\ .
\label{eq:A.22}
\ee
Combining this with Eq.~(\ref{eq:A.21}) yields
\bse
\label{eqs:A.23}
\be
\langle v_p^i \vert p_j\,\epsilon_p\rangle = \delta_{ij} (Ts + n\mu)
\label{eq:A.23a}
\ee
and hence
\be
\langle{\bm v}_p \vert {\bm p}\,\epsilon_p\rangle = 3(Ts+n\mu)\ .
\label{eq:A.23b}
\ee
\ese
Now the Duhem-Gibbs relation, $\mu = G/N$ with $G$ the Gibbs free energy, yields the general identity
\be
\left(\frac{\partial\mu}{\partial T}\right)_{N,V} = \frac{-s}{n} + \frac{1}{n}\left(\frac{\partial p}{\partial T}\right)_{N,V}\ .
\label{eq:A.24}
\ee
This allows us to write Eq.(\ref{eq:A.23a}) as
\be
\langle v_p^i \vert p_j\,\epsilon_p\rangle = \delta_{ij} n \left[ \mu - T\left(\frac{\partial\mu}{\partial T}\right)_{N,V} + \frac{T}{n}\left(\frac{\partial p}{\partial T}\right)_{N,V} \right]\ .
\label{eq:A.25}
\ee
Combining this with Eq.~(\ref{eq:A.7}) yields
\bse
\label{eqs:A.26}
\be
\langle v_p^i\vert p_j\,a_5({\bm p})\rangle = \delta_{ij} T(\partial p/\partial T)_{N,V}
\label{eq:A.26a}
\ee
and hence
\be
\langle {\bm v}_p\vert{\bm p}\,a_5({\bm p})\rangle = 3 T(\partial p/\partial T)_{N,V}
\label{eq:A.26b}
\ee
\ese
From this result we obtain the (2,3) and (3,2) matrix elements in Eq.~(\ref{eq:3.12b}). 

We note that Eqs.~(\ref{eq:A.19a}) and (\ref{eq:A.26b}) are valid for an arbitrary QP velocity
${\bm v}_p$, i.e., and arbitrary interaction function $F({\bm p},{\bm p}')$, not just for the particular form (\ref{eq:2.10a}).

\subsection{The speed of first sound}
\label{app:A.4}

To derive Eq.~(\ref{eq:3.13}) we start with the following expression for the 
ratio $c_p/c_V$ \cite{Landau_Lifshitz_V_1980}
\bse
\label{eqs:A.27}
\be
\frac{c_p}{c_V} = 1 - \frac{T}{V}\,\frac{\left[(\partial V/\partial T)_{p,N}\right]^2}{c_V (\partial V/\partial p)_{T,N}}\ .
\label{eq:A.27a}
\ee
By means of general Jacobian identities we can rewrite this as
\bea
\frac{c_p}{c_V} &=& 1 - \frac{T}{c_V}\frac{1}{V} \left(\frac{\partial V}{\partial p}\right)_{T,N}\left[ \left(\frac{\partial p}{\partial T}\right)_{V,N}\right]^2
\nonumber\\
&=& 1 + \frac{T \chi_T}{c_V} \left[\left(\frac{\partial p}{\partial T}\right)_{V,N}\right]^2\ ,
\label{eq:A.27b}
\eea
\ese
with $\chi_T$ the isothermal compressibility from Eq.~(\ref{eq:A.20b}). This shows that the second
equality in Eq.~(\ref{eq:3.13}) follows form the first one.

\section{Continuity equations}
\label{app:B}

In this appendix we discuss the continuity equations associated with the conservation of the particle number,
the momentum, and the energy, respectively, in the context of our kinetic theory.

\subsection{Particle number conservation}
\label{app:B.1}

Consider the kinetic equation as written in Eqs.~(\ref{eqs:2.16}). Multiplying from the left with the constant function
$\langle a_1({\bm p})\vert = \langle 1\vert$ we obtain
\bse
\label{eqs:B.1}
\be
-iz\,\delta n({\bm k},z) + i{\bm k}\cdot{\bm j}_n({\bm k},z) = \delta n({\bm k},t=0)\ ,
\label{eq:B.1a}
\ee
where the divergence of the number-current density fluctuation is given by
\be
i{\bm k}\cdot{\bm j}_n({\bm k},z) = \langle 1\vert L_{\bm k}^{(1)}({\bm p}) \vert \phi({\bm p},{\bm k},z) \rangle
\label{eq:B.1b}
\ee
\ese
Since the number current density is given by $n\,\delta{\bm u}({\bm k},z)$, with $\delta{\bm u}$ from Eq.~(\ref{eq:2.17b}),
this implies 
\be
 \frac{i k}{m} \langle a_2({\bm p}) \vert \phi({\bm p},{\bm k},z) \rangle = \langle 1 \vert L_{\bm k}^{(1)}({\bm p}) \vert \phi({\bm p},{\bm k},z) \rangle\ .
\label{eq:B.2}
\ee
Since this holds for arbitrary functions $\phi$, we have
\be
\frac{1}{m}\,\hat{\bm k}\cdot{\bm p} = \hat{\bm k}\cdot{\bm v}_p + \frac{1}{N_0 V}\sum_{{\bm p}'} w({\bm p}')\,F({\bm p},{\bm p}')\,(\hat{\bm k}\cdot{\bm p}')\ ,
\label{eq:B.3}
\ee
which is the longitudinal part of Eq.~(\ref{eq:2.7}). Equation~(\ref{eq:B.3}) is valid for any interaction function $F$. With the specific model
interaction given by Eq.~(\ref{eq:2.8}) we obtain the longitudinal part of Eq.~(\ref{eq:2.10a}). We see that Eq.~(\ref{eq:2.10a}) is necessary
for particle-number conservation to hold.

\subsection{Energy conservation}
\label{app:B.2}

The continuity equation related to energy conservation is conveniently expressed in terms of the heat mode,
Eq.~(\ref{eq:3.16}), which is a linear combination of the energy and the density. Multiplying Eq.~(\ref{eq:2.16a}) 
from the left with $\langle \psi_5^{(0)L}({\bm p})\vert$ we find 
\bse
\label{eqs:B.4}
\bea
-i z \left\langle \psi_5^{(0)L}({\bm p})\big\vert \phi({\bm p},{\bm k},z)\right\rangle + i{\bm k}\cdot{\bm j}_T({\bm k},z) &=&
\nonumber\\
&&\hskip -120pt  \left\langle \psi_5^{(0)L}({\bm p})\big\vert \phi({\bm p},{\bm k},t=0)\right\rangle\ .
\label{eq:B.4a}
\eea
The heat current density is given by
\be
{\bm j}_T({\bm k},z) = \left\langle{\bm v}_p \psi_5^{(0)L}({\bm p})\big\vert \phi({\bm p},{\bm k},z)\right\rangle\ ;
\label{eq:B.4b}
\ee
\ese
it determines the heat conductivity via Eq.~(\ref{eq:3.39}). As we discussed in Sec.~\ref{par:III.A.2.b}, the physical 
interpretation of the heat mode is an entropy density. Accordinly, ${\bm j}_T$ is an entropy current density. We note
that this identification of the entropy density as a conserved quantity does not violate the Second Law since
the description of irreversibility requires going past the linearized kinetic theory considered in this paper. 
%$\\$

\subsection{Momentum conservation}
\label{app:B.3}

To obtain the continuity equation related to momentum conservation we multiply Eq.~(\ref{eq:2.16a}) from the
left with $\langle{\bm p}\vert$. This yields 
\begin{widetext}
\be
-iz\, n m\, \delta{\bm u}({\bm k},z) + ik \langle{\bm p}(\hat{\bm k}\cdot{\bm v}_p)\vert\phi({\bm p},{\bm k},z)\rangle 
 + ik\,\frac{F_0}{\langle 1\vert 1\rangle}\,\langle{\bm p}\vert \hat{\bm k}\cdot{\bm v}_p\rangle\, \delta n({\bm k},z) =  nm \,\delta{\bm u}({\bm k},t=0)\ .
\qquad
\label{eq:B.5}
\ee
\end{widetext}
Here we have used the model interaction from Eq.~(\ref{eq:2.8}); this can be generalized if desirable.
By using various identities from Appendix~\ref{app:A}, as well as the identity $-3\vert\sigma_2({\bm p})\rangle = {\cal P}_{\perp} \vert{\bm p}\cdot{\bm v}_p\rangle$
noted after Eq.~(\ref{eq:3.44b}), this can be written
\bse
\label{eqs:B.6}
\be
-iz\, n m\, \delta u^i({\bm k},z) + i k_j \tau^{ij}({\bm k},z) = nm \,\delta u^i ({\bm k},t=0)\ .
\label{eq:B.6a}
\ee
Here 
\be
\tau^{ij}({\bm k},z) = \delta^{ij} \delta p({\bm k},z) + \left\langle \sigma^{ij}({\bm p})\vert \phi({\bm p},{\bm k},z)\right\rangle
\label{eq:B.6b}
\ee
\ese
is the stress tensor. The first term is the reactive part, which is given by the pressure fluctuation $\delta p$, Eq.~(\ref{eq:2.19h}).
The second term is the dissipative part, with $\sigma^{ij}({\bm p})$ from Eqs.~(\ref{eqs:3.43}).

If we break up the momentum into its longitudinal and transverse component, respectively, as in Eqs.~(\ref{eq:2.18})
and (\ref{eqs:2.19}), this can be written
\bse
\bea
-iz\, nm\, \delta u_{\perp}({\bm k,z}) + ik\,j_{u\perp}({\bm k},z) &=& n m\, \delta u_{\perp}({\bm k},t=0)\ ,
\nonumber\\
\label{eq:B.7a}\\
-iz\, nm\, \delta u_{\text{L}}({\bm k,z}) + ik\,j_{u\text{L}}({\bm k},z) &=& n m\, \delta u_{\text{L}}({\bm k},t=0)
\nonumber\\
\label{eq:B.7b}
\eea
for the transverse and longitudinal components, respectively. The transverse momentum current
\be
j_{u\perp}({\bm k},z) =  {\hat k}_{\perp}^i {\hat k}^j \left\langle (\sigma_1)_{ij}({\bm p})\vert \phi({\bm p},{\bm k},z)\right\rangle
\label{eq:B.7c}
\ee
agrees with Eq.~(\ref{eq:3.33b}) and determines the shear viscosity via Eq.~(\ref{eq:3.34}).
The longitudinal one,
\be
j_{u\text{L}}({\bm k},z) = \delta p({\bm k},z) + {\hat k}_i {\hat k}_j \left\langle \sigma^{ij}({\bm p}) \vert \phi({\bm p},{\bm k},z)\right\rangle
\label{eqs:B.7d}
\ee
\ese
combines with the remaining contributions to the sound mode to form Eq.~(\ref{eq:3.42a}).
$\delta u_{\perp}$ and $\hat{\bm k}_{\perp}$ in Eqs.~(\ref{eq:B.7a}, \ref{eq:B.7c}) represent either
of the two transverse directions. 

\smallskip
\section{Linearized collision operators}
\label{app:C}

The linearized collision operator $\Lambda({\bm p})$ is in general a sum of several terms that represent fermion-fermion interactions,
fermion-impurity interactions, and fermion-boson interactions. The latter include, e.g., electron-phonon interactions or electron-magnon interactions
in metals.
%There may be other contributions (e.g., electron-magnon interactions in magnetic metals) that we will not consider here. 
We will denote these three contributions by $\Lambda_{\text{f-f}}$, $\Lambda_{\text{f-b}}$,
and $\Lambda_{\text{f-i}}$, respectively. They can be obtained by linearizing the more general expressions given in
Ref.~\onlinecite{Landau_Lifshitz_X_1981}.

The linearized fermion-fermion collision operator can be written
\begin{widetext}
\bse
\label{eqs:C.1}
\bea
\Lambda_{\text{f-f}}({\bm p})\,\phi({\bm p}) &=& \frac{1}{1-f_{\text{eq}}({\bm p})}\,\frac{1}{V^3} \sum_{{\bm p}',{\bm p}_1,{\bm p}_1'}
     W({\bm p},{\bm p}_1;{\bm p}',{\bm p}_1')\,\delta(\epsilon_p + \epsilon_{p_1} - \epsilon_{p'} - \epsilon_{p_1'})\, \delta({\bm p}+{\bm p}_1-{\bm p}'-{\bm p}_1') 
     \nonumber\\
     && \times f_{\text{eq}}({\bm p}_1) [1 - f_{\text{eq}}({\bm p}')] [1 - f_{\text{eq}}({\bm p}_1')]\,[\phi({\bm p}') + \phi({\bm p}_1') - \phi({\bm p}) - \phi({\bm p}_1)]\ . 
\label{eq:C.1a}
\eea
Here $W$ is the probability for two fermions in momentum states ${\bm p}$ and ${\bm p}_1$ to be scattered into momentum
states ${\bm p}'$ and ${\bm p}_1'$. Time reversal symmetry implies
\be
W({\bm p},{\bm p}_1;{\bm p}',{\bm p}_1') = W({\bm p}',{\bm p}_1';{\bm p},{\bm p}_1)\ .
\label{eq:C.1b}
\ee
\ese
In order to obtain Eq.~(\ref{eq:C.1a}) from the general expression given in Ref.~\onlinecite{Landau_Lifshitz_X_1981} we have
repeatedly used the energy-conservation expressed by the first $\delta$-function as well as the explicit form of the equilibrium
distribution function, Eq.~(\ref{eq:2.3a}). Equation~(\ref{eq:C.1a}) makes explicit the five conservation laws: 
$\Lambda_{\text{f-f}}({\bm p})\,\phi({\bm p}) = 0$ if $\phi({\bm p}) = a_{\alpha}({\bm p})$ with $a_{\alpha}$ any of the five
functions defined in Eqs.~(\ref{eqs:2.20}).

The fermion-impurity collision operator is\cite{impurity_scattering_footnote_1, impurity_scattering_footnote_2}
\be
\Lambda_{\text{f-i}}({\bm p}) \,\phi({\bm p}) =  \frac{1}{V} \sum_{{\bm p'}} W({\bm p}',{\bm p})\,\delta(\epsilon_{p'} - \epsilon_p) \left[\phi({\bm p}') - \phi({\bm p})\right]\ .
\label{eq:C.2}
\ee
Here the fermion particle number and energy are conserved, but the momentum is not.

Finally, the fermion-boson collision operator can be written
\bea
\Lambda_{\text{f-b}}({\bm p})\,\phi({\bm p}) &=&  \frac{1}{f_{\text{eq}}({\bm p})(1-f_{\text{eq}}({\bm p}))}\,\frac{1}{V^2} \sum_{{\bm p}',{\bm k}}
     \delta({\bm p}'-{\bm p}-{\bm k})\,n_{\text{eq}}({\bm k}) \Bigl[W({\bm p}';{\bm p},{\bm k})\,f_{\text{eq}}({\bm p})(1-f_{\text{eq}}({\bm p}'))\,
          \delta(\epsilon_{p'}-\epsilon_p-\omega_k)
\nonumber\\
          && \hskip 100pt + W({\bm p}',-{\bm k};{\bm p})\,f_{\text{eq}}({\bm p}')(1-f_{\text{eq}}({\bm p}))\,\delta(\epsilon_{p'}-\epsilon_p+\omega_k) \Bigr]
               \left[\phi({\bm p}') - \phi({\bm p})\right]\ .
\label{eq:C.3}
\eea
Here $n_{\text{eq}}({\bm k}) = 1/(\exp(\omega_k/T)-1)$ is the equilibrium Bose-Einstein distribution function, and $\omega_k$ is the energy
of a boson with wave number $k$. $W$ is again a transition probability. If the bosons are phonons, then umklapp processes can be 
taken into account by adding a reciprocal lattice vector to the argument of the momentum-conserving $\delta$-function. From Eq.~(\ref{eq:C.3}) 
we see that the fermion particle number is still conserved, but the fermion momentum and energy are not.

For explicit calculations a model fermion-fermion collision operator that is the quantum version of the Bhatnagar-Gross-Krook 
collision operator in classical kinetic theory.\cite{Bhatnagar_Gross_Krook_1954} is also useful. It uses a simple relaxation-time
approximation with a momentum-independent collision rate $1/\tau$ that is temperature dependent. For an ordinary Fermi liquid
it is given, up to a prefactor of $O(1)$, by
\be
1/\tau \approx T^2/\epsilonF
\label{eq:C.2.1}
\ee
with $\epsilonF$ the Fermi temperature. The five conservation laws are taken into account by
projecting on the orthogonal space $\cal{L}_{\perp}$:
\bse
\label{eqs:C.4}
\be
\Lambda_{\text{f-f}}^{\text{BGK}}({\bm p}) = \frac{-1}{\tau}\,{\cal P}_{\perp}\ .
\label{eq:C.4a}
\ee
Here 
\be
{\cal P}_{\perp} = \openone -  \sum_{\alpha=1}^5 \frac{\vert a_{\alpha}({\bm p}\rangle\langle a_{\alpha}({\bm p})\vert}{\langle a_{\alpha}({\bm p})\vert a_{\alpha}({\bm p})\rangle}\ ,
\label{eq:C.4b}
\ee
\ese
with $\openone$ the unit operator and the $a_{\alpha}({\bm p})$ from Eqs.~(\ref{eqs:2.20}),
is the projection operator onto $\cal{L}_{\perp}$ that we have used, e.g., in Eq.~(\ref{eq:3.28a}).
Acting with $\Lambda_{\text{f-f}}^{\text{BGK}}({\bm p})$ on $\phi$ yields
\be
\Lambda_{\text{f-f}}^{\text{BGK}}({\bm p})\,\phi({\bm p},{\bm x},t) = \frac{-1}{\tau} \left[ \phi({\bm p},{\bm x},t) - \frac{1}{N_0}\,\delta n({\bm x},t)
 - \frac{m}{m^*}\,{\bm p}\cdot\delta{\bm u}({\bm x},t) - \frac{1}{Tc_V}\left[\delta e({\bm x},t) - \langle\epsilon_p\rangle_w \delta n({\bm x},t)\right]
     a_5({\bm p}) \right].
\label{eq:C.5}
\ee
\end{widetext}
Substituting Eq.~(\ref{eq:C.5}) in (\ref{eq:2.16b}), and solving the resulting kinetic equation (\ref{eq:2.16a}) in the
hydrodynamic regime, one obtains the same results as in Sec.~\ref{sec:III} (as must be the case), but
with explicit expressions for the transport coefficients and the speed of sound in terms of the parameter
$\tau$. Some of the results have been quoted in the main text.

\acknowledgments

We thank Jim Sauls for a discussion. This work was performed in part at the Aspen Center for Physics, which is supported by 
National Science Foundation grant PHYS-1607611.

%\bibliography{soft_modes_kinetic_theory}

\begin{thebibliography}{55}
\bibitem[{sem()}]{semantics_footnote}
\bibinfo{note}{Strictly speaking one should call these systems `Fermi fluids',
  as there is no phase transition from a Fermi gas to a Fermi liquid, even
  though Landau's original term is universally translated as `Fermi liquid'.}
\bibitem[{the()}]{thermodynamic_limit_footnote}
\bibinfo{note}{Throughout this paper we assume the thermodynamic limit,
  $V\to\infty$, so $(1/V)\sum_{\bm p}$ should be considered as synonymous with
  $\int d{\bm p}/(2\pi)^d$ in $d$ spatial dimensions.}
\bibitem[{v_p()}]{v_p_footnote}
\bibinfo{note}{Within LFL theory, ${\bm v}_p$ is always collinear with ${\bm
  p}$, which follows from Eq.~(\ref{eq:2.7}). However, ${\bm v}_p$ is
  proportional to ${\bm p}$ only if the LFL interaction function $F({\bm
  p},{\bm p}')$ depends only on the angle between ${\bm p}$ and ${\bm p}'$. In
  general, the interaction function also depends on the moduli of the momenta.
  If this is the case, then ${\bm p}\cdot{\bm v}_p$ can no longer be expressed
  as a linear combination of $a_5$ and $a_1$ from Eqs.~(\ref{eqs:2.20}), and
  hence is not an element of the zero-eigenvector space ${\cal L}_0$ defined
  after Eq.~(\ref{eq:2.22}).}
\bibitem[{Lap()}]{Laplace_trafo_footnote}
\bibinfo{note}{Following Refs.~\onlinecite{Forster_1975} and
  \onlinecite{Zubarev_1960}, we define the Laplace transform $f(z)$ of a
  time-dependent function $f(t)$ by $f(z) = \pm \int_{-\infty}^{\infty}
  dt\,\Theta(\pm t) e^{izt}\,f(t)$, with $\pm$ for $\sgn\Im z \gtrless 0$.}
\bibitem[{con()}]{conservation_laws_footnote}
\bibinfo{note}{This is true if one considers a clean closed fermion system.
  Taking into account scattering by phonons, or by impurities, will change the
  number of conservation laws within the fermion system. In general, the
  collision operator will be a sum of various terms that represent different
  scattering processes, see Appendix~\ref{app:C}.}
\bibitem[{bul()}]{bulk_viscosity_footnote}
\bibinfo{note}{In general there are other contributions to the bulk viscosity
  that arise from the momentum dependence of the collision operator.}
\bibitem[{imp({\natexlab{b}})}]{impurity_scattering_footnote_2}
\bibinfo{note}{Strictly speaking the energy-conserving $\delta$-function has a
  finite width since the states with energy $\epsilon_p$ have a finite
  lifetime. A related point is that the QP momentum is not a good quantum
  number, see the discussion Ref.~\onlinecite{Landau_Lifshitz_V_1980}
  $\S\,68$.}
\end{thebibliography}

\end{document}